\documentclass[smallextended,11pt,psfig,a4]{amsart}
\usepackage{latexsym}
\usepackage{amsmath}
\usepackage{amsfonts}
\usepackage{amssymb}

\usepackage{subcaption}
\usepackage{caption}
\usepackage{multirow, bigdelim}
\usepackage{graphicx}

\usepackage{mathrsfs}

\usepackage{geometry}
\usepackage[pdftex]{hyperref}
\usepackage{amssymb}

\newtheorem{theorem}{Theorem}[section]

\newtheorem{pro}[theorem]{Proposition}
\newtheorem*{conj*}{Conjecture}

\newtheorem{remark}[theorem]{Remark}

\theoremstyle{definition}

\newtheorem{example}[theorem]{Example}

\theoremstyle{remark}

\numberwithin{equation}{section}

\newcommand{\ceil}[1]{\lceil {#1}\rceil}

\newcommand{\opn}[1]{\operatorname{#1}}

\newcommand{\mbb}[1]{\mathbb{#1}}

\newcommand{\bs}[1]{\boldsymbol{#1}}

\def\eps{\varepsilon}
\def\>{\rangle}
\def\<{\langle}

\def\Tr{\opn{Tr}}

\def\0{\bs{0}}
\def\1{\mathbbm{1}}

\def\C{\mbb{C}}

  \def\XXint#1#2#3{{\setbox0=\hbox{$#1{#2#3}{\int}$}
      \vcenter{\hbox{$#2#3$}}\kern-.47\wd0}}

\usepackage{tikz}

\tikzstyle{vertex}=[circle, draw, inner sep=0pt, minimum size=4pt]

\usepackage{xcolor,soul}

\usetikzlibrary{arrows,shapes,decorations,automata,backgrounds,petri,positioning}

\tikzset{main node/.style={circle,draw,minimum size=1cm,inner sep=0pt},
}

\begin{document}
\pagestyle{plain}

\def\beq{\begin{equation}}
\def\eeq{\end{equation}}
\def\eps{\epsilon}
\def\laa{\langle}
\def\raa{\rangle}
\def\qed{\begin{flushright} $\square$ \end{flushright}}
\def\qee{\begin{flushright} $\Diamond$ \end{flushright}}
\def\ov{\overline}
\def\bma{\begin{bmatrix}}
\def\ema{\end{bmatrix}}

\def\ora{\overrightarrow}

\def\bma{\begin{bmatrix}}
\def\ema{\end{bmatrix}}
\def\bex{\begin{example}}
\def\eex{\end{example}}
\def\beq{\begin{equation}}
\def\eeq{\end{equation}}
\def\eps{\epsilon}
\def\laa{\langle}
\def\raa{\rangle}
\def\qed{\begin{flushright} $\square$ \end{flushright}}
\def\qee{\begin{flushright} $\Diamond$ \end{flushright}}
\def\ov{\overline}

\author[ag]{Manuel D. de la Iglesia}
\address{Manuel D. de la Iglesia, Instituto de Matem\'aticas,
Universidad Nacional Aut\'onoma de M\'exico. Circuito Exterior, C.U., 04510 Ciudad de M\'exico, M\'exico.}
\email{mdi29@im.unam.mx}

\author[cfelipe]{Carlos F. Lardizabal}
\address{Carlos F. Lardizabal, Instituto de Matem\'atica e Estat\'istica, Universidade Federal do Rio Grande do Sul. Porto Alegre, RS  91509-900 Brazil.}
\email{cfelipe@mat.ufrgs.br}

\date{\today}

\title{One-dimensional Continuous-Time Quantum Markov Chains: qubit probabilities and measures} 

\begin{abstract} 
Quantum Markov chains (QMCs) are positive maps on a trace-class space describing open quantum dynamics on graphs. Such objects have a statistical resemblance with classical random walks, while at the same time they allow for internal (quantum) degrees of freedom. In this work we study continuous-time QMCs on the integer line, half-line and finite segments, so that we are able to obtain exact probability calculations in terms of the associated matrix-valued orthogonal polynomials and measures. The methods employed here are applicable to a wide range of settings, but we will restrict ourselves to classes of examples for which the Lindblad generators are induced by a single positive map, and such that the Stieltjes transforms of the measures and their inverses can be calculated explicitly.
\end{abstract}

\maketitle



\section{Introduction}

In this work, we follow a recent trend coming from quantum information theory, namely, the study of quantum versions of random walks on graphs \cite{salvador}. This is done by taking a classical model as a starting point and investigating statistical notions which have a proper applicability to quantum settings. The discrete-time, coined quantum walk on the line is one of the most popular models regarding unitary evolutions on the integer line, on which various problems and solutions have been discussed. Just as important, the case of continuous-time walks provides invaluable insight regarding the statistics of quantum evolutions described by 1-parameter unitary groups \cite{tamon, portugal}. 

\medskip

Regarding the so-called open (dissipative) quantum versions of random walks, one can also define walks that model both the evolution and the loss of information to the environment, in discrete and continuous-time \cite{attal,sinayskiy4}. In this work we will focus on the latter, and our motivation comes from a class of examples seen in the classical theory of Markov processes. A tridiagonal matrix of the form
\beq\label{eq1}
\mathcal{A}=\begin{bmatrix} -(\mu_0+\lambda_0) & \mu_1 &  &  &  \\
\lambda_0 & -(\mu_1+\lambda_1) & \mu_2 &  &  \\
 & \lambda_1 & -(\mu_2+\lambda_2) & \mu_3 &    \\
 && \ddots & \ddots  & \ddots
 \end{bmatrix},
 \eeq
is the generator, or infinitesimal operator, of a so-called {\bf birth-death process} on the half-line \cite{dIbook}. The probability of finding the particle at some specific vertex can be obtained in terms of polynomials $(Q_n(x))_{n\geq 0}$ defined by the relation
$$
-xQ(x)=Q(x)\mathcal{A},\;\;\;Q(x)=(Q_0(x),Q_1(x),Q_2(x),\dots),\quad Q_0(x)=1,\quad Q_{-1}(x)=0.
$$
Favard's Theorem \cite{chih} guarantees that there exists at least one probability measure $\psi$ supported on the interval $[0,+\infty)$ such that the polynomials defined above are orthogonal with respect to $\psi$. Then, it is well-known that if $P_{ij}(t)$ is the probability of being at vertex $i$ at time $t$, given that the process has started at vertex $j$, we have the {\bf Karlin-McGregor formula} \cite{dIbook,KMc1,KMc2},
\begin{equation*}\label{Kmcontd}
P_{ij}(t)=\frac{\displaystyle\int_{0}^{\infty}e^{-tx}Q_i(x)Q_j(x)d\psi(x)}{\displaystyle\int_{0}^{\infty}Q_i^2(x)d\psi(x)}.
\end{equation*}

In this work, we are interested in open quantum versions of this classical setting. More precisely, we will consider the context of quantum Markov chains (QMCs), as defined by S. Gudder \cite{gudder}, where we have an operator specifying transitions between vertices of any fixed graph. Such transition effects are given by positive maps acting on appropriate spaces of density matrices. In the 1-dimensional cases that we will study in this work, the QMC can be described by block tridiagonal matrices \cite{dette}, generalizing the structure of \eqref{eq1}. For completeness, this setting will be revised in this work. Regarding QMCs and orthogonal polynomials, we recall recent results on such matters:

\medskip

\begin{enumerate}
\item In \cite{jl}, the authors have discussed the problem of finding matrix-valued measures for discrete-time Open Quantum Walks (OQWs)\cite{petr_survey} on the integer half-line. We note that OQWs are a particular case of QMCs for which the transition operators are given by conjugation maps $V_i\cdot V_i^*$, for some matrices $V_i$. In the mentioned work, emphasis was given to the case in which one had commuting transitions.
\item In \cite{dILL}, the authors considered matrix-valued measures for discrete-time QMCs on the integer line, half-line and finite segments. In this setting, the transition operators are assumed to be completely positive (CP) maps, so this setting includes the one of \cite{jl} as a particular case. Non-commuting transitions and non-positive matrix-valued measures were studied in some particular examples. 
\end{enumerate}

We remark that the notion of QMC defined in \cite{gudder} is related, but different from the one presented by L. Accardi \cite{accardi}, which is applied to quantum walks in \cite{accardi2}, see also \cite{dhahri,mukh}, where such kind of construction is applied to OQWs.

\medskip

The present article, on the other hand, besides considering continuous-time dynamics, discusses two directions which were not studied in previous works: 

\medskip

a) {\bf Underlying classical structure.} If the transition operators of a QMC are relatively simple (e.g., corresponding to Hermitian matrices), can we obtain quantum information of the system in terms of some underlying, classical polynomial structure? We will see that under certain conditions, a linear change of coordinates on the components of the QMC will allow us to answer this question positively, and to obtain nonclassical information on the walk. 

\medskip

b) {\bf Transition probabilities for arbitrary sites in practical terms.} It is known that with a quantum version of the Karlin-McGregor formula one can obtain transition probabilities for QMCs. In the works mentioned above, such formula is deduced and explored to some extent, but in practice the calculations seen in \cite{dILL} concerned mostly the case of site recurrence (that is, $i=j$), so that one needs the matrix-valued measure for the walk, but knowing the associated matrix-valued orthogonal polynomials explicitly was not needed. As for the case $i\neq j$, knowledge of the polynomials is essential, and in this work we discuss a proper setting of QMCs where we are able to present closed formulae for arbitrary transition probabilities in terms of Bessel functions. Up to our knowledge, this kind of description has not been explored in the context of OQWs and QMCs so far. Also, we will be able to study site recurrence of these processes.

\medskip

We note that the results presented in this work extend naturally to 1-dimensional QMCs acting on qudits, but we restrict our discussion to chains acting on qubits for the sake of simplicity, as the matrices appearing in the examples are smaller and generalizations are relatively easy to perform if desired.

\medskip

In our presentation, we will focus on the simplest nonclassical instances of QMCs, namely, Hermitian walks of one qubit moving on the integer line, half-line, finite segment, and will present a brief, systematic discussion on such settings. Regarding the integer line, we will consider translation-invariant walks, meaning that we have the same transition rule on every vertex; on the half-line and the finite segment we will consider analogous models, specifying the proper boundary conditions (absorbing or reflecting vertices). In addition, we also make clear that even by restricting the transition rules to be positive operators, the task of obtaining probabilities in terms of a positive definite, matrix-valued measure may be impossible: informally, there should be a basic Hermitian structure underlying the class of examples we will study. Otherwise, the positivity of the measure could be lost, a situation we will refrain from studying in this article, and is the matter of a future work (see also the discussion in the concluding section of this article).

\medskip

Regarding transition probabilities, such calculations play a crucial role in, e.g., the context of localization \cite{balach}, state transfer \cite{christ,godsil, godsil2} and sedentary walks on graphs \cite{godsil3}. However, dissipation is usually absent from such contexts as the evolution is essentially unitary. Therefore, whenever dissipative dynamics is added to the typical unitary development, we believe that the framework of QMCs and associated semigroups may serve as a proper starting point. 

\medskip

The contents of this work are as follows. In Section \ref{sec2}, we describe the basic setting of QMCs and associated continuous-time generators, as well as the QMC version of the Karlin-McGregor formula; we also review the so-called PQ-channels, which are CP maps allowing for a simpler analysis while at the same time produces non-classical statistics. Regarding QMCs induced by channels with Hermitian matrix representations, which is the focus of this work, we also discuss their relevance with respect to the class of all examples such that a positive definite matrix measure is available. In Section \ref{sec4}, we study QMCs on the integer half-line for the case of reflecting and absorbing boundaries. We briefly remark on the case for which we have non-commuting transition operators. Section \ref{sec5} studies QMCs on the integer line. We conclude with Section \ref{sec6}, which describes QMCs on finite segments, and Section \ref{sec7} with a brief discussion on future directions. For the benefit of the reader we also include the Appendix \ref{app1} containing some important formulas related with Chebychev polynomials and Bessel functions which will be used along this paper.

\medskip

\section{The setting: 1-dimensional continuous-time QMCs}\label{sec2}

Let $M_n=M_n(\mathbb{C})$ denote the set of order $n$ matrices with complex coefficients. For $A\in M_n$, we denote by $A^*$ its Hermitian adjoint, and write $A\geq 0$ whenever $A$ is a positive-semidefinite matrix. We recall that a linear map $\Phi:M_n\to M_n$ is {\bf positive} if $\Phi(X)\geq 0$ whenever $X\geq 0$, and we say $\Phi$ is {\bf unital} if $\Phi(I)=I$. We say that $\Phi$ is {\bf trace-preserving} whenever $\mathrm{Tr}(\Phi(X))=\mathrm{Tr}(X)$ for all $X\in M_n$. We say that $\Phi$ is {\bf completely positive (CP)} if there exist matrices $V_i\in M_n$ such that
$$\Phi(X)=\sum_{i=1}^k V_iXV_i^*,\;\;\;\forall\;X\in M_n.$$
This definition is equivalent to the one given in terms of $n$-positive maps, which we will not need in this work. We refer the reader to \cite{bhatia} for more on positive and CP maps on matrix spaces. A CP, trace-preserving map will be called a {\bf quantum channel}.

\medskip

Let $\mathcal{D}_n$ denote the following set of density matrices,
$$\mathcal{D}_n=\Big\{\rho=(\dots,\rho_{-2},\rho_{-1},\rho_0,\rho_1,\rho_2,\dots): \rho_i\in M_n,\; \rho_i\geq 0, \; \sum_{i\in\mathbb{Z}} \mathrm{Tr}(\rho_i)=1\Big\}.$$
Such densities, which consist of the direct sum of positive matrices for which its trace norm equals 1, can also be written in the form
$$\rho\in\mathcal{D}_n\;\Longrightarrow\; \rho=\sum_{i\in\mathbb{Z}}\rho_i\otimes |i\rangle\langle i|,$$
and it is clear that $\mathrm{Tr}(\rho)=\sum_{i\in\mathbb{Z}}\mathrm{Tr}(\rho_i)$. We say that the positive matrix $\rho_i$ corresponds to the information on vertex $i\in\mathbb{Z}$. The number $n$ is the {\bf internal degree of freedom} of the particle. In what follows we still define our setting for arbitrary $n$, but in the remaining sections of this work we will focus on the case $n=2$, that is, on each vertex of the walk we have an order 2 matrix representing some information associated with a qubit.

\medskip

\begin{figure}[h!]
$$\begin{tikzpicture}[>=stealth',shorten >=1pt,auto,node distance=2cm]
\node[state] (q-2)      {$-2$};
\node[state]         (q-1) [right of=q-2]  {$-1$};
\node[state]         (q0) [right of=q-1]  {$0$};
\node[state]         (q1) [right of=q0]  {$1$};
\node[state]         (q2) [right of=q1]  {$2$};
\node         (qr) [right of=q2]  {$\ldots$};
\node         (ql) [left of=q-2]  {$\ldots$};
\path[->]          (ql)  edge         [bend left=15]  node[auto] {$R_{-3}^{-2}$}     (q-2);
\path[->]          (q-2)  edge         [bend left=15]   node[auto] {$R_{-2}^{-1}$}     (q-1);
\path[->]          (q-1)  edge         [bend left=15]   node[auto] {$R_{-1}^{0}$}     (q0);
\path[->]          (q0)  edge         [bend left=15]   node[auto] {$R_0^1$}     (q1);
\path[->]          (q1)  edge         [bend left=15]   node[auto] {$R_{1}^2$}     (q2);
\path[->]          (q2)  edge         [bend left=15]   node[auto] {$R_{2}^3$}     (qr);
\path[->]          (qr)  edge         [bend left=15] node[auto] {$R_{3}^2$}       (q2);
\path[->]          (q2)  edge         [bend left=15]   node[auto] {$R_{2}^1$}     (q1);
\path[->]          (q1)  edge         [bend left=15]  node[auto] {$R_{1}^0$}      (q0);
\path[->]          (q0)  edge         [bend left=15]   node[auto] {$R_{0}^{-1}$}     (q-1);
\path[->]          (q-1)  edge         [bend left=15]  node[auto] {$R_{-1}^{-2}$}      (q-2);
\path[->]          (q-2)  edge         [bend left=15]  node[auto] {$R_{-2}^{-3}$}      (ql);
  \draw [->] (q1) to[loop above]node[auto] {$G_{1}$}  (q1);
   \draw [->] (q2) to[loop above]node[auto] {$G_{2}$}  (q2);
    \draw [->] (q-2) to[loop above] node[auto] {$G_{-2}$} (q-2);
     \draw [->] (q-1) to[loop above]node[auto] {$G_{-1}$}  (q-1);
      \draw [->] (q0) to[loop above]node[auto] {$G_{0}$}  (q0);
\end{tikzpicture}$$
\caption{Nearest neighbor continuous-time OQW $\mathcal{L}$ on $\mathbb{Z}$.}
\label{originalQMConZ}
\end{figure}
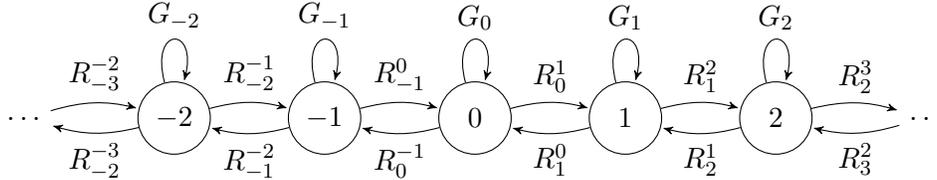

A {\bf nearest neighbor, continuous-time Open Quantum Walk (CTOQW)} is a semigroup $T=(T_t)_{t\geq 0}$ acting on $\mathcal{D}_n$, where
$$
T_t=e^{t\mathcal{L}},\;\;\;t\geq 0,
$$
and $\mathcal{L}$ is the associated {\bf Lindblad generator} \cite{gorini,lindblad} which is defined for every $\rho\in\mathcal{D}_n$ as
\beq\label{lindbladg}
\mathcal{L}(\rho)=\sum_{l\in \mathbb{Z}}\Bigg(G_l\rho_l+\rho_l G_l^*+\sum_{j=l-1}^{l+1} R_j^l\rho_jR_j^{l*}\Bigg)\otimes|l\rangle\langle l|,\;\;\;G_l=-iH_l-\frac{1}{2}\sum_{j=l-1}^{l+1} R_l^{j*}R_l^j,\eeq
where $i=\sqrt{-1}$, for every $j, l$, matrix $R_j^l\in M_n$ is the transition effect from vertex $j$ to vertex $l$, describing the dissipative part of evolution and, for every $l$, $H_l\in M_n$ is a Hermitian matrix, accounting for the unitary part of the dynamics. This particular form of generator ensures that for every $t\geq 0$, $T_t$ is a CP map. Note that in the definition of $\mathcal{L}$, the summation in $j$ accounts for all possible effect transitions that arrive at vertex $l$, and this corresponds to going through the $l$-th row of the matrix representation of $\mathcal{L}$. On the other hand, the expression for $G_l$ ensures that we have the trace-preservation condition
$$G_l+G_l^*+\sum_j R_l^{j*}R_l^j=0,$$
for all $l\in\mathbb{Z}$. This generalizes the classical condition for continuous-time Markov chains, namely, that the {\it columns} of the generator must have entries adding up to 0. Therefore, note that in $G_l$, the indices for $R_l^j$ are inverted with respect to the ones appearing in $\mathcal{L}$. In this work we assume, as in \cite{BBPP}, that $R_l^l=0$ for all $l$ and $R_j^l=0$ if $|j-l|>2$. The underlying graph structure can be seen in Figure \ref{originalQMConZ}.

\medskip

If, in the definition of CTOQW above, we replace the conjugations $R_j^l \cdot R_j^{l*}$ with summations of conjugations, that is, $T_j^l(X)=\sum_{k=1}^{n_{j,l}}R_{j}^{l}(k) X (R_j^{l}(k))^*$ for some $R_j^{l}(k)\in M_n$, with $n_{j,l}$ a finite integer, we call such generalization a {\bf nearest neighbor, continuous-time Quantum Markov Chain (CTQMC)} (or just {\bf QMC} for simplicity) on the integer line, written as
\beq\label{lindbladg0}
\mathcal{L}(\rho)=\sum_{l\in \mathbb{Z}}\Bigg(G_l\rho_l+\rho_l G_l^*+\sum_{j=l-1}^{l+1} T_j^l(\rho_j)\Bigg)\otimes|l\rangle\langle l|,\;\;\;G_l=-iH_l-\frac{1}{2}\sum_{j=l-1}^{l+1}\sum_{k=1}^{n_{l,j}} (R_l^{j}(k))^*R_l^{j}(k).\eeq

We will say that a QMC is {\bf homogeneous} if all of the $T_j^l$ are equal. In the case of CTOQWs this means that $R_j^l=R\in M_n$ for all $j, l\in\mathbb{Z}$. In this work we will deal mostly with homogeneous QMCs, and we refer the reader to Section \ref{sec23} for the case of non-symmetric walks. We note that semigroups given by homogeneous QMCs are trace-preserving, in the sense that $\mathrm{Tr}(e^{t\mathcal{L}}(\rho))=\mathrm{Tr}(\rho)$ for every $t\geq 0$ and $\rho\in \mathcal{D}_n$. QMCs on the half-line and finite segments are defined in an analogous manner. 

\medskip

Regarding the problems studied in this work, with respect to matrices $R_{j}^{l}(k)$ appearing in (\ref{lindbladg0}), we will be mostly interested in the matrix representations of the operators induced by them and inspect whether these are, e.g., Hermitian or positive definite.

\medskip

Now we note that the generator $\mathcal{L}$ can be seen as a linear map acting on $\mathcal{D}_n$, so that the above action can be described in terms of an appropriate tridiagonal block matrix representation. In fact, first we define the vector representation $\mbox{vec}(A)$ of $A\in M_n$, given by stacking together its rows. For instance,
\begin{equation*}
A = \begin{bmatrix} a_{11} & a_{12} \\ a_{21} & a_{22} \end{bmatrix}
\quad\Rightarrow\quad
\mbox{vec}(A):=\begin{bmatrix} a_{11} \\ a_{12} \\ a_{21} \\ a_{22}\end{bmatrix}.
\end{equation*}
The $\mbox{vec}$ mapping establishes a unitary equivalence between the Hilbert spaces $M_n$ and $\C^{n^2}$ with the corresponding standard inner products, $\langle \, \cdot \mid \cdot \, \rangle_{M_n}$ and $\langle \, \cdot \mid \cdot \, \rangle_{\C^{n^2}}$, since
$$
\langle B | A \rangle_{M_n} = \Tr(B^*A) = \sum_{i,j}\overline{B_{ij}}A_{ij} = \mbox{vec}(B)^*\mbox{vec}(A)
= \langle\mbox{vec}(B) | \mbox{vec}(A) \rangle_{\C^{n^2}},
\quad A,B\in M_n.
$$
Also, this mapping satisfies $\mbox{vec}(AXB^T)=(A\otimes B)\,\mbox{vec}(X)$ for any matrices $A, B, X\in M_n$, where $\otimes$ denotes the Kronecker product of matrices \cite{hj2}. In particular, $\mbox{vec}(BXB^*)=\mbox{vec}(BX\ov{B}^T)=(B\otimes \ov{B})\,\mbox{vec}(X)$, and more generally, we have the {\bf matrix representation} of the maps
\begin{equation*}\label{matrep}
T(X):=AXB\;\Longleftrightarrow\; T(X)=\mbox{vec}^{-1}(\ceil{T}\mbox{vec}(X)),\;\;\;\ceil{T}=A\otimes B^T,\;\;\;X\in M_n.
\end{equation*}
Also, for $A\in M_n$, we denote by $\ceil{A}$ the matrix representation of the conjugation $X\mapsto AXA^*$, $X\in M_n$. Therefore, we write $\ceil{A}:=A\otimes\ov{A}$. Now turning to expression \eqref{lindbladg}, we have the following representations,
$$
\rho_l\mapsto G_l\rho_l+\rho_lG_l^*\;\Longleftrightarrow\; \mbox{vec}(\rho_l)\mapsto (G_l\otimes I+I\otimes\ov{G_l})\mbox{vec}(\rho_l),$$
$$\rho_j\mapsto R_j^l\rho_jR_j^{l*} \;\Longleftrightarrow\; \mbox{vec}(\rho_j)\mapsto (R_j^l\otimes \ov{R_j^l})\mbox{vec}(\rho_j),
$$
so the {\bf block matrix representation} for $\mathcal{L}$ can be written as
\begin{equation*}
\widehat{\mathcal{L}}:=\left[\begin{array}{ccc|ccccc}
\ddots & \ddots &  &  &  &  &  &  \\
  \ddots & \mathcal{G}_{-2} & \ceil{R_{-1}^{-2}} &  &  &  &  &  \\
   & \ceil{R_{-2}^{-1}} & \mathcal{G}_{-1} & \ceil{R_{0}^{-1}} &  &  &  &  \\
  \hline
   &    & \ceil{R_{-1}^0} & \mathcal{G}_{0} & \ceil{R_{1}^0}& & &  \\
    &  &    & \ceil{R_{0}^1} & \mathcal{G}_{1} & \ceil{R_{2}^1} & & \\
    & & &    & \ceil{R_{1}^2} & \mathcal{G}_{2} & \ceil{R_{3}^2} &  \\
     &  &  &  &  & \ddots & \ddots &\ddots
\end{array}\right],\;\;\;\mathcal{G}_l:=G_l\otimes I+I\otimes\ov{G_l},\quad l\in\mathbb{Z},
\end{equation*}
which acts on the {\bf vector representation of densities in $\mathcal{D}_n$}, namely,
$$
\overrightarrow{\rho}=\begin{bmatrix} \vdots \\ \mbox{vec}(\rho_{-1}) \\ \mbox{vec}(\rho_0) \\ \mbox{vec}(\rho_1) \\ \vdots\end{bmatrix}.
$$
Hence, we arrive at the differential equation
$$
\frac{d}{dt}\overrightarrow{\rho}(t)=\widehat{\mathcal{L}}\;\overrightarrow{\rho}(t).
$$
The case of QMCs acting on the integer half-line and on finite segments are constructed in a similar way. The above matrix and vector representations are extremely useful regarding the explicit calculation of examples.

\medskip

With respect to the calculation of probabilities, we proceed as follows. Given some initial density matrix $\rho$ concentrated at vertex $j$, written $\rho\otimes|j\rangle\langle j|$, the probability of reaching vertex $i$ at time $t>0$, under the dynamics of $T_t$, is given by 
\begin{equation}\label{pc1}
\mathbb{P}_{j;\rho}^{\,i}(t):=\mathrm{Pr}(\rho\otimes|j\rangle\langle j|\stackrel{t}{\to} |i\rangle\langle i|)=\mathrm{Tr}\Big(\mathbb{Q}_{i}e^{t\mathcal{L}}\rho\otimes |j\rangle\langle j|\Big),\;\;\;\mathbb{Q}_{i}=Q_i\cdot Q_i,\;\;\;Q_i:=I\otimes|i\rangle\langle i|.
\end{equation}
The term inside the trace corresponds to a preparation of an initial density $\rho$, concentrated at vertex $j$, and such state evolves according to the Lindblad generator $\mathcal{L}$. The generator accounts for a quantum operation which usually involves the presence of some kind of dissipation. Projection $\mathbb{Q}_{i}$ means that we are monitoring the arrival  onto vertex $i$ with any internal state and, physically, it corresponds to the measurement of the position of the particle on the graph: such procedure inspects whether the particle can be found at vertex $i$. If the answer is positive, the experiment is over (see \cite{werner} for a description of this monitoring procedure in the discrete time case). If we are interested in the probability of reaching a goal state $\gamma$ at vertex $i$, this can be written as
\beq\label{pc2}
\mathbb{P}_{j;\rho}^{\, i;\gamma}(t):=\mathrm{Pr}(\rho\otimes|j\rangle\langle j|\stackrel{t}{\to} \gamma\otimes|i\rangle\langle i|)=\mathrm{Tr}\Big(\mathbb{Q}_{\gamma,i}e^{t\mathcal{L}}\rho\otimes |j\rangle\langle j|\Big),\;\;\;\mathbb{Q}_{\gamma,i}=Q_{\gamma,i}\cdot Q_{\gamma,i},\;\;\;Q_{\gamma,i}:=|\gamma\rangle\langle\gamma|\otimes|i\rangle\langle i|.
\eeq
With these definitions we can proceed to define the concept of site recurrence. Following \cite{BBPP} (see also \cite{New2}), we say that a vertex $|i\rangle$ is {\bf $\rho$-recurrent} if
\begin{equation}\label{rrecurrent}
\int_0^\infty \mathbb{P}_{i;\rho}^{\,i}(t) dt=\infty.
\end{equation}
When $|i\rangle$ is recurrent for all initial densities $\rho$, we say that $|i\rangle$ is {\bf recurrent}. Otherwise, if there exists an initial density $\rho$ such that the integral in \eqref{rrecurrent} is finite, then we say that $|i\rangle$ is {\bf transient}. We refer the reader to the examples for concrete instances of these calculations, and for the notion of recurrence in the unitary setting, to \cite{werner,krapivsky}.

\subsection{QMC version of the Karlin-McGregor formula}
A QMC version of the Karlin-McGregor formula can be obtained in the same way as for the classical (scalar) case. For completeness, let us briefly recall this simple, but useful reasoning. Let $W$ be a weight matrix, i.e. a $N\times N$ matrix of measures supported in the real line such that $dW(y)- dW(x)\geq 0$ (positive semidefinite) for $x<y$. Define the matrix-valued inner product given by
\begin{equation}\label{innprod}
(P,Q):=\int_\mathbb{R} P^*(x)dW(x)Q(x),
\end{equation}
where $P$ and $Q$ are matrix-valued square integrable functions with respect to $dW$. Let $(Q_n(x))_{n\geq0}$ denote a sequence of matrix-valued orthogonal polynomials with respect to such inner product, with nonsingular leading coefficients. The set of polynomials will be called orthonormal if $\|Q_n\|^2:=(Q_n,Q_n)=I, n\geq0$. It is well-known that any family of matrix-valued orthogonal polynomials satisfies a three-term recurrence relation of the form
\begin{equation}\label{polrec}
-xQ_n(x)=Q_{n+1}(x)A_n+Q_n(x)B_n+Q_{n-1}(x)C_n,\;\;n\geq0,\quad Q_0(x)=I,\quad Q_{-1}(x)=0,
\end{equation}
for certain $A_n, B_n,C_{n+1}, n\geq0,$ square matrices. This gives rise to a block tridiagonal Jacobi matrix of the form
\begin{equation*}\label{tridiag1}
\mathcal{P}=\begin{bmatrix} B_0 & C_1 & & & 0 \\
                    A_0 & B_1 & C_2 &  & & \\
                     & A_1 & B_2 & C_3 & \\
                      0& & \ddots & \ddots & \ddots\end{bmatrix},
\end{equation*}
so that \eqref{polrec} can be written as $-xQ(x)=Q(x)\mathcal{P}$, where $Q(x)=(Q_0(x),Q_1(x),\dots)$. By setting $f(x,t)=Q(x)P(t)$, where $P(t)$ is the transition probability matrix associated with the infinitesimal operator $\mathcal{P}$, and using the Kolmogorov equation (i.e. $P'(t)=\mathcal{P}P(t))$, this satisfies
$$\frac{\partial f(x,t)}{\partial t}=Q(x)P'(t)=Q(x)\mathcal{P}P(t)=-xf(x,t),\;\;\;f(x,0)=Q(x).$$
Therefore, the solution of the previous equations is given by
$$f(x,t)=e^{-xt}Q(x),$$
which implies that 
$$e^{-xt}Q(x)=Q(x)P(t).$$
Together with the orthogonality of $Q_n(x)$ with respect to $W$, we take the inner product \eqref{innprod} in order to obtain an expression for the $(i,j)$ entry of the block matrix $P(t)$. As $\mathcal{P}$ acts on the vectorized expressions for the densities, we apply $\mathrm{vec^{-1}}$ and take the trace in order to obtain (see \eqref{pc1}):
\beq\label{qmc_kmcabc}
\mathbb{P}_{j;\rho}^{\, i}(t)=\mathrm{\mathrm{Tr}}\left(\mathrm{vec}^{-1}\left[\pi_i\left(\int_{\mathbb{R}}e^{-xt}{Q}_i^*(x)dW(x){Q}_j(x)\right)\mathrm{vec}(\rho)\right]\right),\eeq
where $\pi_i=({Q}_i(x),{Q}_i(x))^{-1}$ (see \cite{grcpam,dILL,jl,New2}).

\subsection{PQ-channels}\label{sec3}

Now we briefly review a set of channels for which the calculation of  probabilities are particularly simple. Note that
$$A=\begin{bmatrix} x & y \\ z & w\end{bmatrix}\in M_2\;\Rightarrow\; \ceil{{A}}=A\otimes\ov{A}=\begin{bmatrix} |x|^2 & x\ov{y} & \ov{x}y & |y|^2 \\
x\ov{z} & x\ov{w} & y\ov{z} & y\ov{w}\\
\ov{x}z & \ov{y}z & \ov{x}w & \ov{y}w\\
|z|^2 & z\ov{w} & \ov{z}w & |w|^2\end{bmatrix}.$$
Let us consider order 2 diagonal and anti-diagonal matrices, that is, of the form
$$B=\begin{bmatrix} x & 0 \\ 0 & w\end{bmatrix},\;\;\;D=\begin{bmatrix} 0 & y \\ z & 0\end{bmatrix},$$
so that the matrix representations for their respective conjugations are
$$\ceil{{B}}=B\otimes\ov{B}=\begin{bmatrix} |x|^2 & 0 & 0 & 0 \\
0 & x\ov{w} & 0 & 0\\
0 & 0 & \ov{x}w & 0\\
0 & 0 & 0 & |w|^2\end{bmatrix},\;\;\;\ceil{{D}}=D\otimes\ov{D}=\begin{bmatrix} 0 & 0 & 0 & |y|^2 \\
0 & 0 & y\ov{z} & 0\\
0 & \ov{y}z & 0 & 0\\
|z|^2 &0 & 0 & 0\end{bmatrix}.$$
For the case of order 2 matrices, we will call {\bf PQ-matrices} the set of diagonal/antidiagonal matrices. Regarding quantum channels, we are interested in the set of channels that admits a Kraus decomposition $\sum_i V_i\cdot V_i^*$ for which all matrices $V_i$ are PQ-matrices: these will be called {\bf PQ-channels}. If $\Phi$ is a PQ-channel acting on $M_2$, its matrix representation is of the form
$$\ceil{\Phi}=\sum_i V_i\otimes\ov{V_i}=\begin{bmatrix} p_{11} & 0 & 0 & p_{12}\\ 0 & q_{11} & q_{12} & 0 \\ 0 &\ov{q_{12}} & \ov{q_{11}} & 0 \\ p_{21} & 0 & 0 & p_{22} \end{bmatrix},\;\;\; p_{ij}\in [0,1], \;\;\; p_{11}+p_{21}=1,\;\;\; p_{12}+p_{22}=1, \;\;\; q_{ij}\in\mathbb{C}.$$
Many examples of channels appearing in quantum information theory are of this form: the bit-flip, phase-flip, bit-phase-flip, amplitude damping, depolarizing channel are PQ-channels \cite{cfrr,nielsen}. Certainly not every channel is a PQ-channel, and we note that given a PQ-channel, not every Kraus decomposition consists of PQ-matrices (nevertheless, the matrix representation is independent of such a choice). We also remark that there are homogeneous QMCs covered in this work which are induced by non-PQ-channels.

\medskip

Now, note that by a permutation of the entries, every matrix representation of a PQ-channel on $M_2$ can be written as
\beq\label{pq_f1}
\begin{bmatrix} P & 0 \\ 0 & Q\end{bmatrix}=\begin{bmatrix} p_{11} & p_{12} & 0 & 0 \\ p_{21} & p_{22} & 0 & 0 \\ 0 & 0 & q_{11} & q_{12} \\ 0 & 0 & \ov{q_{12}} & \ov{q_{11}}\end{bmatrix},\eeq
so that such object can be seen as a direct sum of two order 2 matrices. We call such blocks the {\bf $P$ and $Q$ parts}, respectively. Note that $P$ is a column-stochastic matrix. Even though it is a simpler class than the general case, we will see that QMCs induced by PQ-channels allow us to obtain non-classical probability calculations, that is, statistics that cannot be obtained with stochastic matrices alone.

\subsection{QMCs induced by Hermitian channels and their relevance.}\label{sec23} In this work, we are interested in homogeneous open quantum dynamics on the line such that two requirements must be satisfied:
\begin{enumerate}
\item There exists a positive definite matrix-valued measure that describes the statistics of the evolution.
\item The evolution must be given in terms of a valid Lindblad generator.
\end{enumerate}
Regarding the first requirement, there is a well-known criterion for the existence of a matrix-valued measure, in terms of the blocks of the evolution. For completeness, we state it here as follows.\begin{theorem}\label{dette21}(\cite[Theorem 2.1]{dette}) Assume that the matrices $A_n,C_{n+1},n\geq0,$ in the one-step block  tridiagonal transition matrix
\beq\label{tranmat}
\begin{bmatrix} B_0 & C_1 & & & 0 \\
                    A_0 & B_1 & C_2 &  & & \\
                     & A_1 & B_2 & C_3 & \\
                      0& & \ddots & \ddots & \ddots\end{bmatrix},
\eeq
are nonsingular. Then there exists a positive definite matrix $W$ supported on the real line such that the polynomials defined by
\begin{equation*}
xQ_n(x)=Q_{n+1}(x)A_n+Q_n(x)B_n+Q_{n-1}(x)C_n,\;\;n\geq0,\quad Q_0(x)=I,\quad Q_{-1}(x)=0,
\end{equation*}
are orthogonal with respect to the measure $dW(x)$ in the sense of \eqref{innprod} if, and only if,  there exists a sequence of nonsingular matrices $(R_n)_{n\geq0}$ such that
\begin{enumerate}
\item $R_nB_nR_n^{-1} \textrm{ is Hermitian, }\; \forall\; n\geq0$.
\item $R_n^*R_n=\left(A_0^*\cdots A_{n-1}^*\right)^{-1}(R_0^*R_0)C_1\cdots C_n,\;\;\forall\;n\geq1.$
\end{enumerate}
\end{theorem}
The point of the theorem above is the following: the existence of a sequence of nonsingular matrices $(R_n)_{n\geq0}$ satisfying the above conditions is equivalent to being able to symmetrize the block matrix \eqref{tranmat}. In other words, the evolution is similar to a block Hermitian matrix of the form
\beq\label{simetriz}
\begin{bmatrix}N_0 & M_1 & & & 0 \\
                    M_1^* & N_1 & M_2 &  & & \\
                     & M_2^* & N_2 & M_3 & \\
                      0& & \ddots & \ddots & \ddots\end{bmatrix},
\eeq
for some sequences of matrices $(M_n)_{n\geq1}$, and $(N_n)_{n\geq0}$. So, in our setting, it is possible to analyze non-symmetric walks of the form \eqref{tranmat} as long as we are in the conditions of Theorem \ref{dette21}. With this in mind, we note that the generators on which we will be focusing in this work, namely, of the form
\beq\label{this_w}
\Phi=\begin{bmatrix} 0 &  T &  &    \\
T & 0 & T &    \\  & T & 0 & T   \\  & & \ddots & \ddots&\ddots\end{bmatrix},\;\;\;\mathcal{L}=\Phi-I=\begin{bmatrix} -I &  T &  &    \\
T & -I & T &    \\  & T & -I & T   \\
  & & \ddots & \ddots& \ddots\end{bmatrix},
\eeq
are Lindblad generators (as explained in Remark \ref{rk21} below), and clearly the block matrix representations $\widehat{\Phi}$ and $\widehat{\mathcal{L}}$ satisfy Theorem \ref{dette21} whenever $T$ is a channel whose matrix representation $\ceil{T}$ is Hermitian. This is the case if $T$ admits a representation with Hermitian Kraus matrices. Then, a natural question arises: what about generators satisfying (1) and (2) of Theorem \ref{dette21} which are not of the form \eqref{this_w}? In the homogeneous case, \eqref{simetriz} becomes
\begin{equation*}
\begin{bmatrix} N & M & & & 0 \\
                    M & N & M &  & & \\
                     & M & N & M & \\
                      0& & \ddots & \ddots & \ddots\end{bmatrix},
\end{equation*}
for some Hermitian matrices $M, N$, and we can restate our question as: how large is the class of generators satisfying (1) and (2) of Theorem \ref{dette21} that are distinct from \eqref{this_w}? Let us give our answer in two parts:

\medskip

{\bf a) $M$ and $N$ commute.}  If the blocks commute, then they are simultaneously diagonalizable, so we are led to consider the case for which we have $n^2$ copies (where $n^2\times n^2$ is the dimension of the block matrices) of a scalar dynamics of the form
\begin{equation*}
\mathcal{A}_\lambda=\begin{bmatrix} a &  \lambda &  &   \\
\lambda & a & \lambda &    \\  & {\lambda} & a & \lambda   \\
 &  & \ddots & \ddots& \ddots\end{bmatrix},\;\;\;\lambda=\lambda_i, \;\;\;a=a_i,\;\;\;i=1,\ldots,n^2,\;\;\;a_i\in\mathbb{R}.
\end{equation*}
In the case of 1-qubit dynamics ($n=2$) we will have 4 copies of scalar dynamics. By dividing this generator by $a$ and writing $\tilde{\lambda}=-\lambda/a$, we have essentially the dynamics given by the main case $\mathcal{L}=\Phi-I$ in \eqref{this_w}, which will be studied in the following sections.

\medskip

{\bf b) $M$ do not $N$ commute.} In this case, one can resort, if $M$ is positive definite, to \cite[Theorem 3.1]{duran} (this is applied in Remark \ref{ncomm_ex} below). The mentioned theorem follows from a technical discussion on Chebychev matrix polynomials and, while the resulting matrix of measures is still practical to work with, the underlying classical structure is less obvious in this case, due to the non-commutativity of the blocks. On the other hand, the case where the matrix $M$ is not positive definite is the less straightforward situation, as well as the case where $M$ is singular, where the matrix of measures is always \emph{degenerate}, so it cannot have a sequence of matrix-valued orthogonal polynomials (see \cite[Section 4]{duran} for details).

\medskip

As a conclusion, we see that the class of examples given by generators of the form (\ref{this_w}) is perhaps the most tractable one, essentially including the dynamics for which the blocks commute. We believe this class is rich enough to describe nontrivial quantum examples in terms of a classical structure and proper projections onto subspaces of interest. The case of non-commuting blocks may also lead to quantum (and classical) examples associated with positive measures, but usually requires a distinct approach, such as \cite[Theorem 3.1]{duran}. On a distinct, but related direction, if one wishes to consider matrix-valued measures which are not positive definite, much of the basic structure we are considering is lost, and we refer the reader to \cite[Section 9]{dILL} for a discussion in the discrete-time case.

\medskip

A final point worth discussing regards the general structure of the blocks appearing in the main block diagonal. With respect to the structure of the Lindblad generator, the homogeneous case of $G_i$ defined by \eqref{lindbladg} reduces to $G=-iH-R$ for some positive definite matrix $R$ and Hermitian matrix $H$. Then, define $\mathcal{G}=G\otimes I+I\otimes\ov{G}$, and we assume that $\mathcal{G}$ is (similar to) a Hermitian matrix (so that we can use Theorem \ref{dette21}). Then we have:
\begin{pro} 
Let $G=-iH-R$, where $H$ is Hermitian, $R$ is positive definite, and let $\mathcal{G}=G\otimes  I+I\otimes\ov{G}$. We have that $\mathcal{G}$ is Hermitian if, and only if, $H$ is (similar to) a multiple of the identity matrix. As a consequence, if a homogeneous QMC admits a positive definite measure, then the Hamiltonian part of its generator must be a multiple of the identity matrix.
\end{pro}
The proposition above is an important restriction on the structure of any Lindblad generator that has an associated positive definite matrix of measures. Together with Remark \ref{rk21} below, this criterion may be employed whenever one is looking for examples in terms of some fixed set of transition matrices.

\medskip

\begin{remark}\label{rk21} If $\Phi$ is a quantum channel, the proof that $\Phi-I$ is a valid Lindblad generator goes as follows \cite{wolf}: write $\Phi=\sum_j K_j\cdot K_j^*$ and let $\psi(\rho)=\sum_j A_j\rho A_j^*$, where $A_j=K_j-x_jI$ and $x=(x_j)$ is any unit vector. Then, $(\Phi-I)\rho=\psi(\rho)+\kappa\rho+\rho\kappa^*$, where $\kappa=\sum_j \ov{x_j}A_j$. Trace preservation of $\Phi$ means that $\Phi^*(I)=I$ so that $\psi^*(I)+\kappa+\kappa^*=0$. Therefore, the Hermitian part of $\kappa$ is
$$
\frac{\kappa+\kappa^*}{2}=-\frac{\psi^*(I)}{2}.
$$
If the anti-Hermitian part of $\kappa$ is denoted by $B=(\kappa-\kappa^*)/2=-iH$, with $H^*=H$ (so that $B^*=-B)$, we have that
$$
\kappa=-\frac{\psi^*(I)}{2}-iH.
$$
This implies that
$$
(\Phi-I)\rho=i[\rho,H]+\psi(\rho)-\frac{1}{2}\{\psi^*(I),\rho\}_+,
$$
where $\{\cdot,\cdot\}_+$ is the anti-commutator. Therefore, $H$ accounts for the unitary part of the evolution, the CP map $\psi$ describes the dissipative part and, as it is well-known, this is a standard form for a Lindblad generator.
\end{remark}

\section{QMCs on the integer half-line}\label{sec4}

In this section, we will consider Lindblad generators of the form $\mathcal{L}=\Phi-I$, where $\Phi$ is a homogeneous QMC on 1-qubit acting on the integer half-line. This kind of operator $\mathcal{L}$ is a valid Lindblad generator for a continuous-time QMC (by following the procedure seen in Remark \ref{rk21}, one can find the Hermitian and dissipative parts of the generator, if desired).

\subsection{Walks with an absorbing barrier}\label{sec32}
Let $T=\sum_i V_i\cdot V_i^*$ be a 1-qubit CP map satisfying $\sum_i V_i^*V_i=I/2$. In this section and the following ones, we will assume that the matrix representation $\ceil{T}$ is Hermitian.  Being a normal matrix, we have that $\ceil{T}$ is unitarily diagonalizable \cite{hj2}. Define
\beq\label{lind00a}
\Phi=\begin{bmatrix} 0 &  T &  &    \\
T & 0 & T &    \\  & T & 0 & T   \\  & & \ddots & \ddots&\ddots\end{bmatrix},\;\;\;\mathcal{L}=\Phi-I=\begin{bmatrix} -I &  T &  &    \\
T & -I & T &    \\  & T & -I & T   \\
  & & \ddots & \ddots& \ddots\end{bmatrix}.
\eeq
The operator $\Phi$ can be seen as a discrete-time quantum channel acting on $\mathcal{D}_n$ restricted to the integer half-line, and such that vertex $0$ is absorbing, that is, the particle has a nonzero probability of moving to a separate state, say $-1$, from which it cannot leave.  Regarding the order 4 matrix representation of $T$, we write
\begin{equation}\label{matrepT}
\ceil{T}=BD_T B^{*},\;\;\;D_T=\mbox{diag}(\lambda_1,\lambda_2,\lambda_3,\lambda_4).
\end{equation}
This diagonal form, obtained via the unitary change of coordinates $B$, means that the walk can be regarded as a direct sum of four scalar dynamics. That is, if $\mathcal{R}=\mbox{diag}(B,B,\dots)$, we can write the block matrix representation of $\Phi$ as
$$
\widehat{\Phi}=\mathcal{R}\begin{bmatrix} 0 &  D_T & &    \\
D_T & 0 & D_T &    \\  & D_T & 0 & D_T   \\ & & \ddots & \ddots& \ddots\end{bmatrix}\mathcal{R}^{*},
$$
and similarly for $\widehat{\mathcal{L}}$. As each block $D_T$ is diagonal, we have four scalar evolutions, one for each eigenvalue of $D_T$. Indeed, since 
$$
\mathcal{R}^*\widehat{\mathcal{L}}\mathcal{R}=\begin{bmatrix} -I &  D_T & &    \\
D_T & -I & D_T &    \\  & D_T & -I & D_T   \\ & & \ddots & \ddots& \ddots\end{bmatrix},
$$
then, in order to find the matrix-valued measure associated with the QMC induced by $\mathcal{L}$, we need to study the spectral analysis of the following tridiagonal, or Jacobi matrices:
\beq\label{lind00b}
\mathcal{A}_\lambda=\begin{bmatrix} -1 &  \lambda &  &   \\
\lambda & -1 & \lambda &    \\  & {\lambda} & -1 & \lambda   \\
 &  & \ddots & \ddots& \ddots\end{bmatrix},\;\;\;\lambda=\lambda_i,\;\;\;i=1,2,3,4,
\eeq
each one for every eigenvalue of $\ceil{T}$. In order to do that, one of the standard ways is to find the relation between the spectral measure $\psi_\lambda$ associated with $\mathcal{A}_\lambda$ and the spectral measure $\psi_\lambda^{(0)}$ associated with $\mathcal{A}_\lambda^{(0)}$, which is built from $\mathcal{A}_\lambda$ by removing the first row and column (see \cite[Theorem 1.5]{dIbook}). Let us call $B(z;\omega)=\int (x-z)^{-1}d\omega(x)$ the {\bf Stieltjes transform} associated with any real measure $\omega$. In this case, from \eqref{lind00b}, we see that 
$\mathcal{A}_\lambda=\mathcal{A}_\lambda^{(0)}$ and therefore
$\psi_\lambda=\psi_\lambda^{(0)}$. Hence, using \cite[(1.22)]{dIbook}, we have
$$
B(z;\psi_\lambda)=\frac{-1+z-\sqrt{(z-\sigma_-)(z-\sigma_+)}}{2\lambda^2},\quad \sigma_{\pm}=1\pm2\lambda.
$$
For simplicity, we will assume that $\lambda>0$ in which case we have $\sigma_-<\sigma_+$. In the case of $\lambda<0$, then $\sigma_+<\sigma_-$ and all the computations are similar. Using the Perron-Stieltjes inversion formula (see for instance \cite[Proposition 1.1]{dIbook}), we have that the spectral measure is given by
\beq\label{the_measure}
\boxed{
\psi_\lambda(x)=\frac{\sqrt{(x-\sigma_-)(\sigma_+-x)}}{2\pi\lambda^2},\;\;\;x\in[\sigma_-,\sigma_+]\;\;\;\text{(half-line, absorbing barrier)}}
\eeq
Also, it is possible to compute the orthogonal polynomials via the relation $-xQ^\lambda(x)=Q^\lambda(x)\mathcal{A}_\lambda $, where $Q^\lambda(x)=(Q_0^\lambda(x), Q_1^\lambda(x),\ldots)$ with $Q_0^\lambda(x)=1$. The components of $Q^\lambda(x)$ satisfy a three-term recurrence relation depending on $x$ and has as solution 
$$
Q_n^\lambda(x)=U_n\left(\frac{1-x}{2\lambda}\right),\quad n\geq0,
$$
where $(U_n)_n$ are the Chebychev polynomials of the second kind defined by \eqref{CP2}. The polynomials $(Q_n^\lambda)_n$ are in fact orthonormal with respect to the spectral measure \eqref{the_measure}. One consequence of the above is that we can give a solution of the differential equation $P'(t)= P(t)\mathcal{A}_\lambda, P(0)=I,$ in terms of the spectral measure and the corresponding orthogonal polynomials 
via the classical Karlin-McGregor formula. Indeed, if we call $P^\lambda(t)=(P_{ij}^\lambda(t))$, then we have 
$$
P_{ij}^\lambda(t)=\int_{\sigma_-}^{\sigma_+} e^{-xt}U_i\left(\frac{1-x}{2\lambda}\right)U_j\left(\frac{1-x}{2\lambda}\right)\frac{\sqrt{(x-\sigma_-)(\sigma_+-x)}}{2\pi\lambda^2}dx.
$$
In this case, this formula can be significantly simplified via the change of variables $x=1-2\lambda\cos\theta$. Indeed,
\begin{align}
\nonumber P_{ij}^\lambda(t)&=\int_{0}^{\pi} e^{-(1-2\lambda\cos\theta)t}U_i(\cos\theta)U_j(\cos\theta)\frac{\sqrt{4\lambda^2-4\lambda^2\cos^2\theta}}{2\pi\lambda^2}2\lambda\sin\theta d\theta\\
\label{the_kmg} &=\frac{2e^{-t}}{\pi}\int_0^\pi e^{2\lambda t\cos\theta}\sin[(i+1)\theta]\sin[(j+1)\theta]d\theta\\
\nonumber &=e^{-t}\left(I_{i-j}(2\lambda t)-I_{i+j+2}(2\lambda t)\right),
\end{align}
where $I_\nu(z)$ denotes the modified Bessel function of the first kind defined by \eqref{mBf1}. In the last step we have used standard trigonometric formulas and \eqref{mBf1IR}. 

\medskip

Regarding the matrix-valued measure for $\mathcal{L}$, we write
$$
dW(x)=B\Psi(x)B^{*}dx,\;\;\;\Psi(x)=\mbox{diag}(\psi_{\lambda_1}(x),\psi_{\lambda_2}(x),\psi_{\lambda_3}(x),\psi_{\lambda_4}(x)).
$$
The matrix-valued polynomials can be obtained by writing 
$$\mathcal{Q}_i(x)=BD_i(x)B^{*},$$
$$
D_i(x)=\mbox{diag}(Q_i^{\lambda_1}(x),Q_i^{\lambda_2}(x),Q_i^{\lambda_3}(x),Q_i^{\lambda_4}(x)),\;\;\;Q_i^{\lambda_k}(x)=U_i((1-x)/2\lambda_k),\;\;\;\;k=1,2,3,4,
$$
where $\lambda_i$ are the eigenvalues of $\ceil{T}$. Since $(Q_i^\lambda)_n$ are orthonormal, then $\pi_i=I$ and we have, by the Karlin-McGregor formula for QMCs \eqref{qmc_kmcabc}, that
\begin{equation}\label{tqmc_calc}
\mathbb{P}_{j;\rho}^{\, i}(t)=\mathrm{\mathrm{Tr}}\left(\mathrm{vec}^{-1}\left[B\Big[\int_{\mathbb{R}}e^{-xt}D_i^*(x)D_j(x)\Psi(x)dx\Big]B^{*}\mathrm{vec}(\rho)\right]\right).
\end{equation}
Observe that we have an entry-wise integral calculation over a diagonal matrix. From \eqref{tqmc_calc} for $i=j=0$ one expects that, as $t\to\infty$, such probability will tend to $0$ since the process is absorbing. This forces the support of each of the measures in \eqref{the_measure} to be contained in the interval $[0,\infty)$. Otherwise, such integral would be divergent. This is equivalent to say that $\sigma_-=1-2|\lambda|\geq0$, that is, $|\lambda|\leq 1/2$. Note that,  with respect to \eqref{lind00a}, this is consistent with the assumption that $\sum_i V_i^*V_i=I/2$. Another direct way to conclude this is from \eqref{the_kmg} and the asymptotic formula \eqref{mBf1HE} for the modified Bessel function of the first kind. In this case we have 
$$
e^{-t}I_\nu(2\lambda t)\to\frac{e^{-(1-2\lambda)t}}{\sqrt{4\pi\lambda t}}\left(1+\mathcal{O}(1/t)\right),\quad t\to\infty.
$$
Then, as long as $|\lambda|\leq 1/2$, we have that $e^{-t}I_\nu(2\lambda t)\to0$.



\bex\label{Ex4.1}

Consider a QMC on the half-line as described by \eqref{lind00a}, where $T$ is a 1-qubit PQ-channel divided by 2, with  matrix representation of the form
$$
\ceil{T}=\frac{1}{2}\begin{bmatrix} p_1 & 0 & 0 & p_2 \\ 0 & q & r & 0 \\ 0 & \ov{r} & q & 0 \\ 1-p_1 & 0 & 0 & 1-p_2\end{bmatrix},\;\;\;p_1, p_2\in[0,1],\quad q,r\in\mathbb{C}.
$$
Note that the $P$ part of this map will be normal if and only if $p_2=1-p_1$ and, in this case, $T$ will be unital.  Restricting to this case, we write $p=p_1$ and we have that
the $P$ and $Q$ parts of $T$ are
$$
P=\frac{1}{2}\begin{bmatrix} p & 1-p \\ 1-p & p\end{bmatrix},\;\;\;Q=\frac{1}{2}\begin{bmatrix} q & r \\ r & q\end{bmatrix}.
$$
For the sake of simplicity we consider the case of PQ-channels such that $q, r$ are real, but the situation for which these are complex can be studied in a similar manner. 

\medskip

The associated matrix-valued measure for the QMC with generator $\Phi$ will be the direct sum of the measure for the $P$ part with the one for the $Q$ part, associated with the representation \eqref{pq_f1}, as follows.

\medskip

{\bf Measure for the $Q$ part:} a calculation gives that
$$Q=BCB^{*},\;\;\;B=\frac{1}{\sqrt{2}}\begin{bmatrix} 1 & -1 \\ 1 & 1\end{bmatrix},\;\;\;C=\frac{1}{2}\begin{bmatrix} q+r & 0 \\ 0 & q-r\end{bmatrix},$$
so with respect to (\ref{the_measure}) and \eqref{the_kmg}, we will consider $\lambda_{3}=(q+r)/2$ and $\lambda_{4}=(q-r)/2$. Then, we have
$$\Psi_Q(x)=\frac{2}{\pi}\begin{bmatrix} \frac{\sqrt{(q+r)^2-(x-1)^2}}{(q+r)^2} & 0 \\ 0 & \frac{\sqrt{(q-r)^2-(x-1)^2}}{(q-r)^2}   \end{bmatrix}.$$
Therefore,
$$dW_Q(x)=B\Psi_Q(x)B^{*}=\frac{[\sqrt{(q+r)^2-(x-1)^2}]_+}{\pi(q+r)^2}\begin{bmatrix} 1 & 1 \\ 1 & 1 \end{bmatrix}+\frac{[\sqrt{(q-r)^2-(x-1)^2}]_+}{\pi(q-r)^2}\begin{bmatrix} 1 & -1 \\ -1 & 1\end{bmatrix},$$
where $|x-1|<|q+r|$ for the first measure and $|x-1|<|q-r|$ for the second one. Here we are using the notation $[f (x)]_+ = f (x)$ if $f (x)\geq0$ and 0 otherwise. The condition on $|\lambda|\leq 1/2$ restricts our parameters $q$ and $r$ to be in the following range: $|q+r|\leq1$ and $|q-r|\leq1$.

\medskip

{\bf Measure for the $P$ part:} as $P$ is bistochastic, this is a particular case of the calculation for the $Q$ part: just set $q=p$ and $r=1-p$. Then,
$$P=BDB^{*},\;\;\;B=\frac{1}{\sqrt{2}}\begin{bmatrix} 1 & -1 \\ 1 & 1\end{bmatrix},\;\;\;D=\begin{bmatrix} \frac{1}{2} & 0 \\ 0 & \frac{2p-1}{2}\end{bmatrix},\;\;\;\Psi_P(x)=\frac{2}{\pi}\begin{bmatrix} \sqrt{1-(x-1)^2} & 0 \\ 0 & \frac{\sqrt{4(2p-1)^2-(x-1)^2}}{(2p-1)^2}   \end{bmatrix},$$
$$dW_P(x)=B\Psi_P(x) B^{*}=\frac{[\sqrt{1-(x-1)^2}]_+}{\pi}\begin{bmatrix} 1 & 1 \\ 1 & 1\end{bmatrix} +\frac{[\sqrt{(2p-1)^2-(x-1)^2}]_+}{\pi(2p-1)^2}\begin{bmatrix} 1 & -1 \\ -1 & 1\end{bmatrix},$$
where $|x-1|<1$ for the first measure and $|x-1|<|2p-1|$ for the second one. Above, we write $\lambda_{1}=1/2$ and $\lambda_2=p-1/2$. The condition on $|\lambda|\leq 1/2$ restricts our parameter $p$ to be in the range $0\leq p\leq1$.

\medskip

The above measures were presented for completeness of exposition but, in practice, we can resort directly to \eqref{the_kmg} in order to obtain probabilities. If we call 
$$
\hat{B}=B\oplus B=\begin{bmatrix} B & 0 \\ 0 & B\end{bmatrix},\;\;\;\Psi(x)=\begin{bmatrix} \Psi_P(x) & 0 \\ 0 & \Psi_Q(x)\end{bmatrix},
$$
then, using \eqref{tqmc_calc}, we have that the probability of reaching site $i$ at time $t$, with any internal state at such site, given an initial density $\rho=\frac{1}{2}\begin{bmatrix} 1+X & Y+iZ \\ Y-iZ & 1-X\end{bmatrix}$ at vertex $j$, equals
\begin{equation}\label{probcomp}
\begin{split}
\mathbb{P}_{j;\rho}^{\, i}(t)&=\mathrm{\mathrm{Tr}}\left(\mathrm{vec}^{-1}\left[\pi_j \hat{B}\Big[\int_{\mathbb{R}}e^{-xt}D_i^*(x)D_j(x)\Psi(x)\Big]\hat{B}^{*}\mathrm{vec}(\rho)\right]\right)\\
&=\mathrm{\mathrm{Tr}}\left(\mathrm{vec}^{-1}\left[\begin{bmatrix} \frac{P_{ij}^{\lambda_1}(t)+P_{ij}^{\lambda_2}(t)}{4} & 0 & 0 & \frac{P_{ij}^{\lambda_1}(t)-P_{ij}^{\lambda_2}(t)}{4} \\ 0 & \frac{P_{ij}^{\lambda_3}(t)+P_{ij}^{\lambda_4}(t)}{4} & \frac{P_{ij}^{\lambda_3}(t)-P_{ij}^{\lambda_4}(t)}{4} & 0\\0 & \frac{P_{ij}^{\lambda_3}(t)-P_{ij}^{\lambda_4}(t)}{4} & \frac{P_{ij}^{\lambda_3}(t)+P_{ij}^{\lambda_4}(t)}{4}& 0 \\ \frac{P_{ij}^{\lambda_1}(t)-P_{ij}^{\lambda_2}(t)}{4} & 0 & 0 & \frac{P_{ij}^{\lambda_1}(t)+P_{ij}^{\lambda_2}(t)}{4}\end{bmatrix}\begin{bmatrix} 1+X \\ Y+iZ \\ Y-iZ \\ 1-X\end{bmatrix}\right]\right)\\
&=\frac{1}{4}\Bigg[(P_{ij}^{\lambda_1}(t)+P_{ij}^{\lambda_2}(t))(1+X)+(P_{ij}^{\lambda_1}(t)-P_{ij}^{\lambda_2}(t))(1-X)+\\
&\hspace{2cm}(P_{ij}^{\lambda_1}(t)-P_{ij}^{\lambda_2}(t))(1+X)+(P_{ij}^{\lambda_1}(t)+P_{ij}^{\lambda_2}(t))(1-X)\Bigg]\\
&=P_{ij}^{\lambda_1}(t)=e^{-t}\left(I_{i-j}(t)-I_{i+j+2}(t)\right),\quad i,j\in\mathbb{N}_0,
\end{split}
\end{equation}
recalling that  $\lambda_1=1/2$. From this, we see that with respect to a generator induced by a PQ-channel as constructed above, the probability of reaching a {\it site} is actually a classical calculation, independent of the inner details of the PQ-channel and of the initial density as well. As this example regards PQ-matrices, this is not surprising: the information extracted by a trace calculation on PQ-matrices is given by their $P$-parts, which acts only on the diagonals of density matrices, and a projection onto a site acts as the identity on such site, so that no information coming from the off-diagonal entries of the density is obtained.

Regarding the site recurrence of the process we have to evaluate the integral in \eqref{rrecurrent}. But this integral can be explicitly computed using \eqref{mBf1LT} (for $\alpha=1$ and $s\to1$). Indeed, after some computations as $s\to1$, we obtain
\begin{equation*}\label{recuifin}
\int_0^\infty\mathbb{P}_{i;\rho}^{\,i}(t)dt=\int_0^\infty P_{ii}^{1/2}(t)dt=\int_0^\infty e^{-t}\left(I_{0}(t)-I_{2i+2}(t)\right)dt=2i+2,\quad i\in\mathbb{N}_0.
\end{equation*}
Since this integral is finite for every initial density $\rho$ we have that all vertices are transient, as expected, since we have an absorbing barrier at vertex $0$.

\medskip

On the other hand, the situation is different if one seeks non-classical statistics, e.g. the probability of reaching a {\it specific state} $\psi$ on that site. In this case, one needs to multiply the term inside the trace by the projection onto $\psi$, as prescribed in \eqref{pc2}, and this usually leads to a nontrivial dependence on the channel and on the initial density. For instance, if
$$
|\psi\rangle=\begin{bmatrix}\ov{\psi}_1 \\ \ov{\psi}_2 \end{bmatrix},\;\;\; \gamma=|\psi\rangle\langle\psi|=\begin{bmatrix} |\psi_1|^2 & \ov{\psi_1}\psi_2 \\ \ov{\psi_2}\psi_1 & 1-|\psi_1|^2\end{bmatrix},\;\;\;\Gamma=\gamma\cdot\gamma ,\;\;\;\rho=\frac{1}{2}\begin{bmatrix} 1+X & Y+iZ \\ Y-iZ & 1-X\end{bmatrix},$$
then, recalling that $\pi_i=I$ and $\ceil{\Gamma}=\gamma\otimes \gamma$, we have
\begin{align*}
\Gamma&\hat{B}\Big[\int_{\mathbb{R}}e^{-xt}D_i^*(x)D_j(x)\Psi(x)\Big]\hat{B}^{*}\mathrm{vec}(\rho)=\frac{1}{4}\begin{bmatrix} |\psi_1|^4 & |\psi_1|^2\psi_1\ov{\psi_2} & |\psi_1|^2\psi_2\ov{\psi_1} & |\psi_1|^2|\psi_2|^2 \\
|\psi_1|^2\psi_2\ov{\psi_1}& |\psi_1|^2|\psi_2|^2 & \ov{\psi_1}^2\psi_2^{2} & |\psi_2|^2\psi_2\ov{\psi_1} \\ 
|\psi_1|^2\psi_1\ov{\psi_2} & \ov{\psi_2}^2\psi_1^{2} & |\psi_1|^2|\psi_2|^2 & |\psi_2|^2\psi_1\ov{\psi_2} \\ 
 |\psi_1|^2|\psi_2|^2& |\psi_2|^2\psi_1\ov{\psi_2} & |\psi_2|^2\psi_2\ov{\psi_1} & |\psi_2|^4\end{bmatrix}\times\\
 &\hspace{1cm}\times\begin{bmatrix} P_{ij}^{\lambda_1}(t)+P_{ij}^{\lambda_2}(t) & 0 & 0 & P_{ij}^{\lambda_1}(t)-P_{ij}^{\lambda_2}(t) \\ 0 & P_{ij}^{\lambda_3}(t)+P_{ij}^{\lambda_4}(t) & P_{ij}^{\lambda_3}(t)-P_{ij}^{\lambda_4}(t) & 0\\0 & P_{ij}^{\lambda_3}(t)-P_{ij}^{\lambda_4}(t) & P_{ij}^{\lambda_3}(t)+P_{ij}^{\lambda_4}(t)& 0 \\ P_{ij}^{\lambda_1}(t)-P_{ij}^{\lambda_2}(t) & 0 & 0 & P_{ij}^{\lambda_1}(t)+P_{ij}^{\lambda_2}(t)\end{bmatrix}\begin{bmatrix} 1+X \\ Y+iZ \\ Y-iZ \\ 1-X\end{bmatrix}
\end{align*}
so that
\begin{equation}\label{probrhopsi}
\begin{split}
\mathbb{P}_{j;\rho}^{\, i;\gamma}(t)&=\mathrm{\mathrm{Tr}}\left(\;\mathrm{vec}^{-1}\left[\Gamma\hat{B}\Big[\int_{\mathbb{R}}e^{-xt}D_i^*(x)D_j(x)\Psi(x)\Big]\hat{B}^{*}\mathrm{vec}(\rho)\right]\right)\\
&=X(|\psi_1|^2-1/2)P_{ij}^{\lambda_2}(t)+Y\Re(\ov{\psi_1}\psi_2)P_{ij}^{\lambda_3}(t)+Z\Im(\ov{\psi_1}\psi_2)P_{ij}^{\lambda_4}(t)+\frac{1}{2}P_{ij}^{\lambda_1}(t).
\end{split}
\end{equation}
The above expression clearly shows the dependence on the initial state $\rho$ and the final state $\gamma=|\psi\rangle\langle\psi|$, as well as the initial site $j$ and final site $i$. One natural question arises in this situation: for a fixed final pure state given by $|\psi\rangle=(\psi_1, \psi_2)^*$, is there any choice of the initial state $\rho$ such that the plot of $\mathbb{P}_{j;\rho}^{\, i;\gamma}(t), t\geq0,$ is highest/lowest? To answer this question we need to solve a nonlinear optimization problem. Indeed, observe that $\mathbb{P}_{j;\rho}^{\, i;\gamma}(t)$ can be written as
$$
\mathbb{P}_{j;\rho}^{\, i;\gamma}(t)=a_{ij}^{\lambda_2}(t;\psi)X+b_{ij}^{\lambda_3}(t;\psi)Y+c_{ij}^{\lambda_4}(t;\psi)Z+d_{ij}^{\lambda_1}(t;\psi),
$$
where
\begin{equation}\label{abcd}
\begin{split}
a_{ij}^{\lambda_2}(t;\psi)&=(|\psi_1|^2-1/2)P_{ij}^{\lambda_2}(t),\\
b_{ij}^{\lambda_3}(t;\psi)&=\Re(\ov{\psi_1}\psi_2)P_{ij}^{\lambda_3}(t),\\
c_{ij}^{\lambda_4}(t;\psi)&=\Im(\ov{\psi_1}\psi_2)P_{ij}^{\lambda_4}(t),\\
d_{ij}^{\lambda_1}(t;\psi)&=\frac{1}{2}P_{ij}^{\lambda_1}(t).
\end{split}
\end{equation}
On the other hand, the values of $X,Y,Z$ such that $\rho$ is a density are restricted to the points satisfying  $X^2+Y^2+Z^2\leq 1$. Therefore, we need to optimize $\mathbb{P}_{j;\rho}^{\, i;\gamma}(t)$ subject to $(X,Y,Z)$ satisfying such inequality.  This problem can be easily solved using standard techniques of nonlinear optimization problems like Lagrange multipliers or the Karush-Kuhn-Tucker conditions \cite{rusz}. The optimal solutions are given by 
\begin{equation}\label{optsols}
\begin{split}
X_{\pm}&=\pm\frac{a_{ij}^{\lambda_2}(t;\psi)}{\sqrt{\left[a_{ij}^{\lambda_2}(t;\psi)\right]^2+\left[b_{ij}^{\lambda_3}(t;\psi)\right]^2+\left[c_{ij}^{\lambda_4}(t;\psi)\right]^2}},\\
Y_{\pm}&=\pm\frac{b_{ij}^{\lambda_3}(t;\psi)}{\sqrt{\left[a_{ij}^{\lambda_2}(t;\psi)\right]^2+\left[b_{ij}^{\lambda_3}(t;\psi)\right]^2+\left[c_{ij}^{\lambda_4}(t;\psi)\right]^2}},\\
Z_{\pm}&=\pm\frac{c_{ij}^{\lambda_4}(t;\psi)}{\sqrt{\left[a_{ij}^{\lambda_2}(t;\psi)\right]^2+\left[b_{ij}^{\lambda_3}(t;\psi)\right]^2+\left[c_{ij}^{\lambda_4}(t;\psi)\right]^2}}.
\end{split}
\end{equation}
Observe that these solutions have a high dependence on the sites $i,j$ and the time $t$, as well as the final given state given by $\psi$, so the position of the initial states $\rho^{\pm}=\frac{1}{2}\begin{bmatrix} 1+X_{\pm} & Y_{\pm}+iZ_{\pm} \\ Y_{\pm}-iZ_{\pm} & 1-X_{\pm}\end{bmatrix}$ may depend on $i$, $j$, $t$ and $\psi$.

\medskip

For a concrete instance, take $p=1-s/2$, $q=1-s$ and $r=0$ for $0\leq s\leq1$, which corresponds to a \emph{depolarizing channel}. This implies that $\lambda_1=\frac{1}{2}$ and $\lambda_2=\lambda_3=\lambda_4=\frac{1-s}{2}$. Therefore, from \eqref{probrhopsi} we obtain
$$
\mathbb{P}_{j;\rho}^{\, i;\gamma}(t)=\frac{1}{2}P_{ij}^{1/2}(t)+\left[(|\psi_1|^2-1/2)X+\Re(\ov{\psi_1}\psi_2)Y+\Im(\ov{\psi_1}\psi_2)Z\right]P_{ij}^{\frac{1-s}{2}}(t).
$$
Since $\lambda_2=\lambda_3=\lambda_4=\frac{1-s}{2}$, an easy computation shows that (see \eqref{abcd})
\begin{align*}
\left[a_{ij}^{\lambda_2}(t;\psi)\right]^2&+\left[b_{ij}^{\lambda_3}(t;\psi)\right]^2+\left[c_{ij}^{\lambda_4}(t;\psi)\right]^2=\left[(|\psi_1|^2-1/2)^2+\Re(\ov{\psi_1}\psi_2)^2+\Im(\ov{\psi_1}\psi_2)^2\right]\left[P_{ij}^{\lambda_2}(t)\right]^2\\
&=\left[(|\psi_1|^2-1/2)^2+|\psi_1|^2|\psi_2|^2\right]\left[P_{ij}^{\lambda_2}(t)\right]^2=\left[\frac{1}{2}P_{ij}^{\lambda_2}(t)\right]^2,
\end{align*}
so, from \eqref{optsols}, we obtain that the optimal solutions in the case of the depolarizing channel are independent of $i, j$ and $t$. Indeed,
$$
\rho^+=\begin{bmatrix} |\psi_1|^2 & \ov{\psi_1}\psi_2 \\ \ov{\psi_2}\psi_1 & 1-|\psi_1|^2\end{bmatrix},\quad \rho^-=\begin{bmatrix} 1-|\psi_1|^2 & -\ov{\psi_1}\psi_2 \\ -\ov{\psi_2}\psi_1 & |\psi_1|^2\end{bmatrix}.
$$
The initial state $\rho^+$ corresponds to the final state $|\psi\rangle\langle\psi|$ while the initial state $\rho^-$ is the corresponding antipodal point determined by $[-\ov{\psi_2},  \; \ov{\psi_1}]^*$. This independence on $i, j$ and $t$ only applies in this case of the depolarizing channel, since $\lambda_2=\lambda_3=\lambda_4=\frac{1-s}{2}$. In general $\rho^{\pm}$ will have a strong dependence on $i, j$ and $t$.

\begin{figure}[!ht]
    \centering
    \begin{subfigure}[b]{0.5\textwidth}
        \centering
        \includegraphics[height=3.10in]{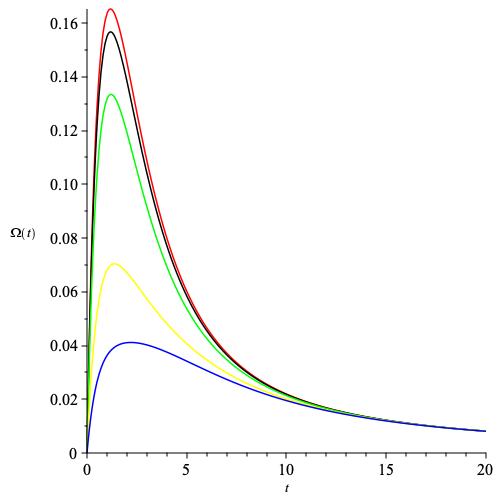}
    \end{subfigure}%
    ~ 
    \begin{subfigure}[b]{0.5\textwidth}
        \centering
        \includegraphics[height=3.10in]{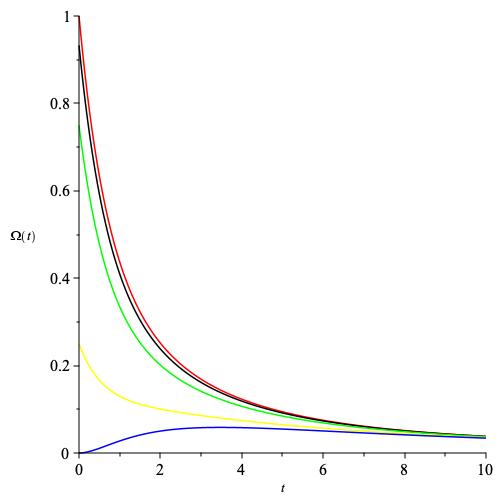}
           \end{subfigure}
  \caption{Plots for the probabilities $\mathbb{P}_{j;\rho}^{\, i;\gamma}(t)$ in \eqref{probrhopsi} regarding a QMC on the half-line with absorbing boundary at $|0\rangle$ with final state $|\psi\rangle=(\frac{1}{2},\frac{\sqrt{3}}{2})^*$, and $p=5/6$, $q=2/3$ and $r=0$, corresponding to a copy of a depolarizing channel on each site. The diagram on the left corresponds to the case of $i=1,j=0,$ while the diagram on the right is for $i=1,j=1$. The highest curve (in red) corresponds to the initial state $\rho^+$, the lowest curve (in blue) corresponds to the initial state $\rho^-$, and the other three curves (from highest to lowest) correspond to the initial states $\rho=\frac{1}{2}\begin{bmatrix} 1 & 1 \\ 1 & 1\end{bmatrix}$ (in black), $\rho=E_{22}$ (in green) and $\rho=E_{11}$ (in yellow).}
\label{fig1}
\end{figure}

For instance, if we take $s=1/3$, we get $p=5/6$, $q=2/3$ and $r=0$. This implies that $\lambda_1=1/2$ and $\lambda_{2}=\lambda_{3}=\lambda_{4}=1/3$. Fixing as a final state $|\psi\rangle=(\frac{1}{2},\frac{\sqrt{3}}{2})^*$, we obtain the diagrams shown in Figure \ref{fig1}, where the highest curve (in red) corresponds to the initial state $\rho^+$, the lowest curve (in blue) corresponds to the initial state $\rho^-$, and the other three curves (from highest to lowest) correspond to the initial states $\rho=\frac{1}{2}\begin{bmatrix} 1 & 1 \\ 1 & 1\end{bmatrix}$ (in black), $\rho=E_{22}:=\begin{bmatrix} 0 & 0 \\ 0 & 1\end{bmatrix}$ (in green) and $\rho=E_{11}:=\begin{bmatrix} 1 & 0 \\ 0 & 0\end{bmatrix}$ (in yellow).

\begin{remark}
From the diagram on the right of Figure \ref{fig1}, regarding the case with equal initial and goal sites, we see that probabilities at time $t=0$ may vary from 0 to 1. For instance, if we take $\rho=\rho^+$, then this probability is obviously 1, but if we take the antipodal point $\rho=\rho^-$, then this probability is 0, since it corresponds to an orthogonal state. In general, the probabilities at $t=0$ and $i=j$ will be nonzero whenever there is an overlap between the initial and goal states, and are given by (see \eqref{probrhopsi})
$$
\mathbb{P}_{i;\rho}^{\, i;\gamma}(0)=(|\psi_1|^2-1/2)X+\Re(\ov{\psi_1}\psi_2)Y+\Im(\ov{\psi_1}\psi_2)Z+1/2.
$$
If $\rho=E_{11}$ then $\mathbb{P}_{i;\rho}^{\, i;\gamma}(0)=|\psi_1|^2$, if $\rho=E_{22}$ then $\mathbb{P}_{i;\rho}^{\, i;\gamma}(0)=1-|\psi_1|^2$ and if $\rho=\frac{1}{2}\begin{bmatrix} 1 & 1 \\ 1 & 1\end{bmatrix}$ then $\mathbb{P}_{i;\rho}^{\, i;\gamma}(0)=1/2+\Re(\ov{\psi_1}\psi_2)$, as we can see in Figure \ref{fig1}.
\end{remark}

\eex
\qee

\subsection{Walks with a reflecting barrier}\label{sec42}

In this case, the matrix \eqref{lind00b} should be replaced by 
\begin{equation*}
\mathcal{A}_\lambda=\begin{bmatrix} \lambda-1 &  \lambda &  &   \\
\lambda & -1 & \lambda &   \\  & \lambda & -1 & \lambda  \\
& & \ddots & \ddots& \ddots\end{bmatrix}, \;\;\;\lambda=\lambda_i,\;\;\;i=1,\dots,4.
\end{equation*}
As before, assuming that $\lambda>0$, we can compute the Stieltjes transform of the spectral measure $\psi_\lambda$ of $\mathcal{A}_\lambda$, given in this case by
$$
B(z;\psi_\lambda)=\frac{z-\sigma_--\sqrt{(z-\sigma_-)(z-\sigma_+)}}{2\lambda(z-\sigma_-)},\quad \sigma_{\pm}=1\pm2\lambda.
$$
Therefore, using the Perron-Stieltjes formula, we have that the spectral measure is given by
\begin{equation*}
\boxed{\psi_\lambda(x)=\frac{1}{2\pi\lambda}\sqrt{\frac{\sigma_+-x}{x-\sigma_-}},\;\;\; x\in(\sigma_-,\sigma_+)\;\;\;\text{(half-line, reflecting barrier)}}
\end{equation*}
In this case, the orthonormal polynomials generated by the three-term recurrence relation $-xQ^\lambda(x)=Q^\lambda(x)\mathcal{A}_\lambda$ with $Q^\lambda(x)=(Q_0^\lambda(x), Q_1^\lambda(x),\ldots)$, are given by
\beq\label{pol1}
Q_n^\lambda(x)=V_n\left(\frac{1-x}{2\lambda}\right),\quad n\geq0,
\eeq
where $(V_n)_n$ are the Chebychev polynomials of the third kind defined by \eqref{CP3}. As before, we can give a solution of the differential equation $P'(t)=P(t)\mathcal{A}_\lambda, P(0)=I,$ in terms of the classical Karlin-McGregor formula, and also can be simplified after the change of variables $x=1-2\lambda\cos\theta$. Indeed,
\begin{align*}
P_{ij}^\lambda(t)&=\frac{1}{2\pi\lambda}\int_{1-2\lambda}^{1+2\lambda} e^{-xt}V_i\left(\frac{1-x}{2\lambda}\right)V_j\left(\frac{1-x}{2\lambda}\right)\sqrt{\frac{1+2\lambda-x}{2\lambda-1+x}}dx\\
&=\frac{e^{-t}}{\pi}\int_{0}^{\pi} e^{2\lambda t\cos\theta}\frac{\cos\left[(i+1/2)\theta\right]\cos\left[(j+1/2)\theta\right]}{\cos^2(\theta/2)}\sqrt{\frac{1+\cos\theta}{1-\cos\theta}}\sin\theta d\theta\\
&=\frac{2e^{-t}}{\pi}\int_{0}^{\pi} e^{2\lambda t\cos\theta}\cos\left[(i+1/2)\theta\right]\cos\left[(j+1/2)\theta\right]d\theta\\
&=\frac{e^{-t}}{\pi}\int_{0}^{\pi} e^{2\lambda t\cos\theta}\left[\cos[(i-j)\theta]+\cos[(i+j+1)\theta)]\right]d\theta\\
&=e^{-t}\left(I_{i-j}(2\lambda t)+I_{i+j+1}(2\lambda t)\right),
\end{align*}
where $I_\nu(z)$ denotes the modified Bessel function of the first kind defined by \eqref{mBf1}. Above we have used \eqref{CP3}, standard trigonometric formulas and, as before, \eqref{mBf1IR}. 

\begin{figure}[!ht]
    \centering
    \begin{subfigure}[b]{0.5\textwidth}
        \centering
        \includegraphics[height=3.10in]{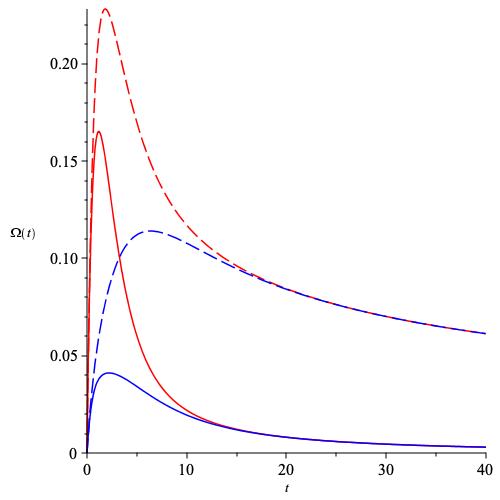}
    \end{subfigure}%
    ~ 
    \begin{subfigure}[b]{0.5\textwidth}
        \centering
        \includegraphics[height=3.10in]{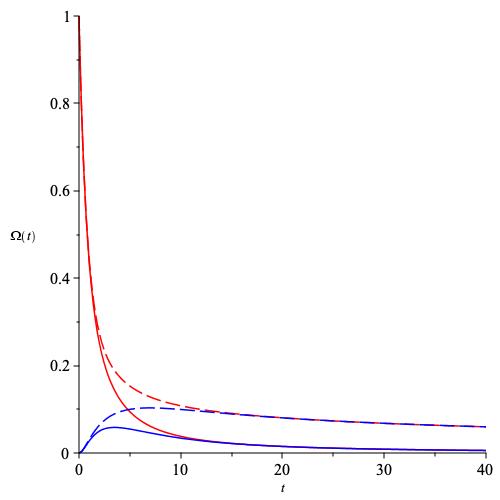}
           \end{subfigure}
  \caption{Comparison between the probability plots of $\mathbb{P}_{j;\rho}^{\, i;\gamma}(t)$ in \eqref{probrhopsi} regarding a QMC on the half-line with absorbing (solid lines) and reflecting (dash lines) boundaries at $|0\rangle$. The values of $\psi, p, q, r, i$ and $j$ are the same as in Figure \ref{fig1}. The highest curves (in red) correspond to the initial state $\rho^+$ and the lowest curves (in blue) correspond to the initial state $\rho^-$.}
\label{fig2}
\end{figure}

In this setting, Example \ref{Ex4.1} can be revisited, obtaining probabilities which are quite similar for small values of $t$ but the reflecting case produces probabilities which decay to $0$ slower than the absorbing case, see Figure \ref{fig2}.
Performing the same computations as in \eqref{probcomp}, we obtain
$$
\mathbb{P}_{j;\rho}^{\, i}(t)=e^{-t}\left(I_{i-j}(t)+I_{i+j+1}(t)\right),\quad i,j\in\mathbb{N}_0.
$$
Therefore, regarding the site recurrence of the process we have to evaluate again the integral in \eqref{rrecurrent}. Now we have that $\mathbb{P}_{i;\rho}^{\, i}(t)=e^{-t}\left(I_{0}(t)+I_{2i+1}(t)\right)$. Using again \eqref{mBf1LT} (for $\alpha=s=1$) we have now that $\int_0^\infty\mathbb{P}_{i;\rho}^{\,i}(t)dt=\infty$ for every initial density $\rho$. Therefore we have that all vertices are recurrent.

\begin{remark}
From \eqref{probrhopsi} one may wonder if the behavior of the integral $\int_0^\infty\mathbb{P}_{i;\rho}^{\, i;\gamma}(t)dt$ is different from the one already studied for the site recurrence. Or, in other words, if there exists a density $\gamma$ such that $\int_0^\infty\mathbb{P}_{i;\rho}^{\, i;\gamma}(t)dt<\infty$ knowing that $\int_0^\infty\mathbb{P}_{i;\rho}^{\,i}(t)dt=\infty$ for all $i\in\mathbb{N}_0$. In this case we have, using \eqref{mBf1LT} (for $\alpha=2\lambda$ and $s=1$), that
$$
\int_0^\infty P_{ii}^{\lambda}(t)dt=\int_0^\infty e^{-t}\left(I_{0}(2\lambda t)+I_{2i+1}(2\lambda t)\right)dt=\frac{1}{\sqrt{1-4\lambda^2}}\left[1+\left(\frac{\sqrt{1-4\lambda^2}-1}{2\lambda}\right)^{2i+1}\right].
$$
From this expression we see that the only value for which this integral is divergent is for $\lambda^2=1/4$. Since the eigenvalues here are given by $\lambda_1=1/2$, $\lambda_2=p-1/2, 0\leq p\leq1$, and $\lambda_3=\frac{q+r}{2},\lambda_4=\frac{q-r}{2}$, for $|q\pm r|\leq1$, we need to find a choice of the parameters such that the divergence generated by the element $P_{ii}^{1/2}(t)$ can be cancelled out somehow. This can be done, for instance, if we choose $r=1-q$, so that $\lambda_3=1/2$.  By taking $\psi_1=1/\sqrt{2}, \psi_2=-1/\sqrt{2}$, so that $\gamma=\frac{1}{2}\begin{bmatrix} 1 & -1 \\ -1 & 1\end{bmatrix}$, and  $\rho=\frac{1}{2}\begin{bmatrix} 1 & 1 \\ 1 & 1\end{bmatrix}$, we have, from \eqref{probrhopsi}, that $\mathbb{P}_{i;\rho}^{\, i;\gamma}(t)=0$. However, for this class of examples, we have not been able to find a situation where $\mathbb{P}_{i;\rho}^{\, i;\gamma}(t)>0$ and $\int_0^\infty\mathbb{P}_{i;\rho}^{\, i;\gamma}(t)dt<\infty$.
\end{remark}

\begin{remark}{\bf(Non-commutative transitions).}\label{ncomm_ex} One can also find classes of examples for which the transitions do not commute. To give the reader an idea regarding how to proceed in this case, let us examine a generator of the form
\begin{equation*}\label{lind0}
\widehat{\mathcal{L}}=\begin{bmatrix} \mathcal{G} &  \ceil{T} & &  \\
\ceil{T} & \mathcal{G} & \ceil{T} &   \\  & \ceil{T} & \mathcal{G} & \ceil{T}  \\
&& \ddots & \ddots& \ddots\end{bmatrix},
\end{equation*}
where vertex $0$ is seen as having a positive probability of absorbing the particle (this property can be modified later if needed, see e.g. \cite[Theorem 2.6]{dette}). Let
$$T=V_1\cdot V_1^*+V_2\cdot V_2^*,\;\;\;V_1=\begin{bmatrix} a & 0 \\ 0 & b\end{bmatrix},\;\;\;V_2=\begin{bmatrix} 0 & c \\ d & 0 \end{bmatrix},\;\;\;\;a,b,c,d\in\mathbb{C}\setminus\{0\}.$$
Then,
$$\ceil{T}=\begin{bmatrix} |a|^2 & 0 & 0 & |c|^2 \\ 0 & a\ov{b} & c\ov{d} & 0 \\ 0 & \ov{c}d & \ov{a}{b} & 0 \\ |d|^2 & 0 & 0 & |b|^2 \end{bmatrix},\;\;\;G=-(V_1^*V_1+V_2^*V_2)=-\begin{bmatrix} |a|^2+|d|^2 & 0 \\ 0 & |b|^2+|c|^2\end{bmatrix},$$
$$\mathcal{G}=G\otimes I_2+I_2\otimes\ov{G}=-\begin{bmatrix} 2(|a|^2+|d|^2) & 0 & 0 & 0 \\ 0 & |a|^2+|b|^2+|c|^2+|d|^2 & 0 & 0 \\ 0 & 0 & |a|^2+|b|^2+|c|^2+|d|^2 & 0 \\ 0 & 0 & 0 & 2(|b|^2+|c|^2)\end{bmatrix},$$
and it holds that $G+G^*+2(V_1^*V_1+V_2^*V_2)=0$. We have that
$$
\ceil{T}\mathcal{G}-\mathcal{G}\ceil{T}=2(|a|^2+|d|^2-|b|^2-|c|^2)\begin{bmatrix} 0 & 0 & 0 & |c|^2\\ 0 & 0 & 0 & 0  \\ 0 & 0 & 0 & 0 \\ -|d|^2 & 0 & 0 & 0  \end{bmatrix},
$$
so $\ceil{T}$ and $\mathcal{G}$ do not commute in general. Nevertheless, there are situations where a positive definite matrix-valued measure can still be obtained. If, for instance, we assume $|c|=|d|$, $|a|\neq |b|$, one can easily find examples for which $\ceil{T}$ is represented by a positive definite matrix and, as $\mathcal{G}$ is a Hermitian matrix, we can resort to Duran's result \cite[Theorem 3.1]{duran}, which considers the discrete-time case, but can be immediately applied to continuous-time QMCs. Indeed, by writing
\begin{equation*}
H(z)=\ceil{T}^{-1/2}(\mathcal{G}-zI)\ceil{T}^{-1}(\mathcal{G}-zI)\ceil{T}^{-1/2}-4I,
\end{equation*}
the associated matrix-valued measure becomes
\begin{equation*}
dW(x)=\frac{1}{2\pi}\ceil{T}^{-1/2}U(x)(D^+(x))^{1/2}U^{*}(x)\ceil{T}^{-1/2}dx,
\end{equation*}
where $D(x)=(d_{ii}(x))$ is diagonal, $D^+(x)$ is the diagonal matrix with entries $d_{ii}^+(x)=\max\{d_{ii}(x),0\}$ and $U(x)D(x)U^*(x)$ is a unitary diagonalization of $-H(x)$, which is Hermitian if $x$ is real. The matrix-valued orthogonal polynomials can be obtained via the usual recurrence relations although, in principle, one does not have a more direct approach to this matter, as the blocks do not commute. Also, and more directly, the state recurrence problem can be examined as well.
\end{remark}

\section{QMCs on the integer line}\label{sec5}

Let $T=\sum_i V_i\cdot V_i^*$ be a 1-qubit CP map satisfying $\sum_i V_i^*V_i=I/2$. Let us define now the discrete-time quantum channel $\Phi$, acting on $\mathcal{D}_n$, as the doubly-infinite tridiagonal block matrix
\begin{equation*}\label{lind00Z}
\Phi=\left[\begin{array}{ccc|ccccc}
\ddots & \ddots &  &  &  &  &  &  \\
  \ddots & 0 & T &  &  &  &  &  \\
   & T &0& T&  &  &  &  \\
  \hline
   &    & T & 0 & T & & &  \\
    &  &    & T & 0 & T & & \\
       &  &  &  & \ddots & \ddots &\ddots
\end{array}\right].
\end{equation*}
As before, using the same notation of the matrix representation for $T$ (see \eqref{matrepT}) we get a diagonal form of $\widehat{\Phi}$, where again we have 4 scalar evolutions, one for each eigenvalue of $D_T$. Now we have to study the spectral analysis of the following doubly-infinite Jacobi matrix
\begin{equation}\label{lind00bZ}
\mathcal{A}_\lambda=\left[\begin{array}{ccc|ccccc}
\ddots & \ddots &  &  &  &  &  &  \\
  \ddots & -1 & \lambda &  &  &  &  &  \\
   & \lambda &-1& \lambda&  &  &  &  \\
  \hline
   &    & \lambda & -1 & \lambda & & &  \\
    &  &    & \lambda & -1 & \lambda & & \\
       &  &  &  & \ddots & \ddots &\ddots
\end{array}\right],\;\;\;\lambda=\lambda_i,\;\;\;i=1,2,3,4.
\end{equation}
In this situation, we need to apply the spectral theorem \emph{three times}, so there will be three spectral measures $\psi_\lambda^{\alpha,\beta},\alpha,\beta=1,2$ (with $\psi_\lambda^{12}=\psi_\lambda^{21}$) which are usually written in compact form as what is called the \emph{spectral matrix}
$$
\Phi_\lambda(x)=\begin{bmatrix}\psi_\lambda^{11}(x) & \psi_\lambda^{12}(x) \\ \psi_\lambda^{12}(x) & \psi_\lambda^{22}(x)\end{bmatrix}.
$$
In order to compute this spectral matrix we also have tools related with the Stieltjes transform of $\Phi$ (see for instance \cite[Section 3.8]{dIbook}). After some computations, we get that the spectral matrix associated with \eqref{lind00bZ} is given by
\begin{equation}\label{measZ}
\boxed{\Phi_\lambda(x)=\frac{1}{\sqrt{(x-\sigma_-)(\sigma_+-x)}}\begin{bmatrix} 1&\displaystyle\frac{1-x}{2\lambda} \\ \displaystyle\frac{1-x}{2\lambda} & 1\end{bmatrix},\quad x\in[\sigma_-,\sigma_+],\quad \sigma_{\pm}=1\pm2\lambda\quad \mbox{(integer line)}}
\end{equation}
Again, it is possible to compute the left-eigenvectors of $\mathcal{A}_\lambda$, which are now given by two families of linearly independent polynomials given by the relation $-xQ^{\lambda,\alpha}(x)=Q^{\lambda,\alpha}(x)\mathcal{A}_\lambda$, $\alpha=1,2$, where $Q^{\lambda,\alpha}(x)=(\ldots, Q_{-1}^{\lambda,\alpha}(x), Q_0^{\lambda,\alpha}(x), Q_1^{\lambda,\alpha}(x),\ldots)$ and the initial conditions $Q_{-1}^{\lambda,1}(x)=0$, $Q_{0}^{\lambda,1}(x)=1$ and $Q_{-1}^{\lambda,2}(x)=1$, $Q_{0}^{\lambda,2}(x)=0$. The components of $Q^{\lambda,\alpha}(x)$, $\alpha=1,2$ are now given by 
\begin{equation*}
Q_n^{\lambda,1}(x)=\begin{cases}
U_n\left(\displaystyle\frac{1-x}{2\lambda}\right),&\mbox{if} \quad n\geq0,\vspace{.3cm}\\
-U_{-n-2}\left(\displaystyle\frac{1-x}{2\lambda}\right),&\mbox{if} \quad n\leq-1,\\
\end{cases}
\end{equation*}
and
\begin{equation*}
Q_n^{\lambda,2}(x)=\begin{cases}
-U_{n-1}\left(\displaystyle\frac{1-x}{2\lambda}\right),&\mbox{if} \quad n\geq0,\vspace{.3cm}\\
U_{-n-1}\left(\displaystyle\frac{1-x}{2\lambda}\right),&\mbox{if} \quad n\leq-1,\\
\end{cases}
\end{equation*}
where $(U_n)_n$ are the Chebychev polynomials of the second kind defined by \eqref{CP2}. These two families of polynomials are orthonormal with respect to the spectral matrix \eqref{measZ} in the following sense
$$
\int_{\mathbb{R}}\left[Q_n^{\lambda,1}(x),Q_n^{\lambda,2}(x)\right]\Phi_\lambda(x)\begin{bmatrix}Q_m^{\lambda,1}(x)\\Q_m^{\lambda,2}(x)\end{bmatrix}dx=\sum_{\alpha,\beta=1}^2\int_{\mathbb{R}}Q_n^{\lambda,\alpha}(x)\psi_\lambda^{\alpha,\beta}(x)Q_m^{\lambda,\beta}(x)dx=\delta_{n,m},\quad n,m\in\mathbb{Z}.
$$
As before we can give a solution of the differential equation $P'(t)=P(t)\mathcal{A}_\lambda, P(0)=I,$ in terms of the spectral matrix and the corresponding two families of orthogonal polynomials 
via the Karlin-McGregor formula. Indeed, if we call $P^\lambda(t)=(P_{ij}^\lambda(t))_{i,j\in\mathbb{Z}}$, then we have, for $i,j\geq0$ (similar computations if $i,j\leq -1$, $i\geq0,j\leq-1$ and $i\leq-1,j\geq0$), and after the change of variables $x=1-2\lambda\cos\theta$, that
\begin{align}
\nonumber P_{ij}^\lambda(t)&=\frac{e^{-t}}{\pi}\int_{0}^{\pi}e^{2t\lambda\cos\theta}\left[U_i\left(\cos\theta\right),-U_{i-1}\left(\cos\theta\right)\right]\begin{bmatrix} 1&\cos\theta \\ \cos\theta& 1\end{bmatrix}\begin{bmatrix}U_j\left(\cos\theta\right)\\-U_{j-1}\left(\cos\theta\right)\end{bmatrix}d\theta\\
\nonumber &=\frac{e^{-t}}{\pi}\int_{0}^{\pi}e^{2t\lambda\cos\theta}\left[U_i\left(\cos\theta\right)-\cos\theta U_{i-1}\left(\cos\theta\right),\cos\theta U_i\left(\cos\theta\right)-U_{i-1}\left(\cos\theta\right)\right]\begin{bmatrix}U_j\left(\cos\theta\right)\\-U_{j-1}\left(\cos\theta\right)\end{bmatrix}d\theta\\
\nonumber &=\frac{e^{-t}}{\pi}\int_{0}^{\pi}e^{2t\lambda\cos\theta}\left[T_i\left(\cos\theta\right),T_{i+1}\left(\cos\theta\right)\right]\begin{bmatrix}U_j\left(\cos\theta\right)\\-U_{j-1}\left(\cos\theta\right)\end{bmatrix}d\theta\\
\nonumber &=\frac{e^{-t}}{\pi}\int_{0}^{\pi}e^{2t\lambda\cos\theta}\left[T_i\left(\cos\theta\right)U_j\left(\cos\theta\right)-T_{i+1}\left(\cos\theta\right)U_{j-1}\left(\cos\theta\right)\right]d\theta\\
\label{kmgZ}&=\frac{e^{-t}}{\pi}\int_{0}^{\pi}e^{2t\lambda\cos\theta}T_{i-j}\left(\cos\theta\right)d\theta=e^{-t}I_{i-j}(2\lambda t),
\end{align}
where $(T_n)_n$ are the Chebychev polynomials of the first kind defined by \eqref{CP1} and $I_\nu(z)$ is the modified Bessel function of the first kind defined by \eqref{mBf1}. Once again we have used \eqref{mBf1IR} and previously \eqref{cp1cp2}.

\medskip

\begin{figure}[!ht]
    \centering
    \begin{subfigure}[b]{0.5\textwidth}
        \centering
        \includegraphics[height=3.10in]{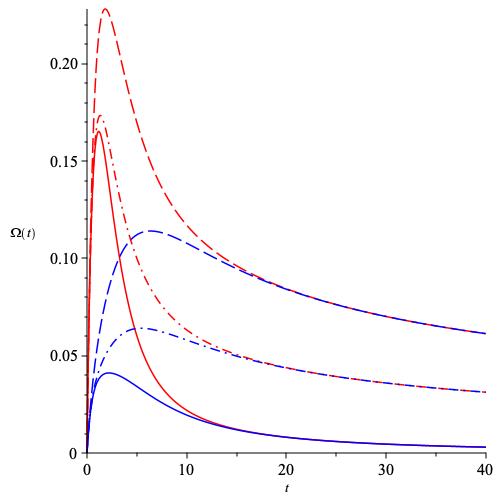}
    \end{subfigure}%
    ~ 
    \begin{subfigure}[b]{0.5\textwidth}
        \centering
        \includegraphics[height=3.10in]{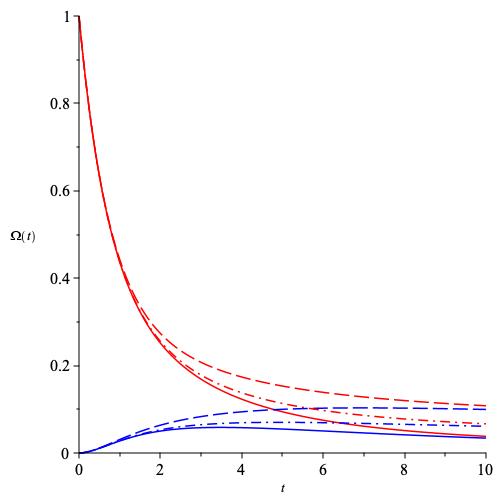}
           \end{subfigure}
  \caption{Comparison between the probability plots of $\mathbb{P}_{j;\rho}^{\, i;\gamma}(t)$ in \eqref{probrhopsi} regarding a QMC on the integer line (dash-dots lines), the absorbing half-line (solid lines) and the reflecting half-line (dash lines). The values of $\psi, p, q, r, i$ and $j$ are the same as in Figure \ref{fig1}. The highest curves (in red) correspond to the initial state $\rho^+$ and the lowest curves (in blue) correspond to the initial state $\rho^-$.}
\label{fig3}
\end{figure}

Regarding the matrix-valued measure for $\mathcal{L}$ (see \eqref{matrepT}), we write
$$
dW_{\alpha,\beta}(x)=B\Psi_{\alpha,\beta}(x)B^{*}dx,\;\;\;\Psi_{\alpha,\beta}(x)=\mbox{diag}(\psi_{\lambda_1}^{\alpha,\beta}(x),\psi_{\lambda_2}^{\alpha,\beta}(x),\psi_{\lambda_3}^{\alpha,\beta}(x),\psi_{\lambda_4}^{\alpha,\beta}(x)),\quad \alpha,\beta=1,2.
$$
We can derive a Karlin-McGregor formula for this continuous-time 
QMC version in a similar way as we did in \cite[Section 8]{dILL}. Indeed, we have
$$
\mathbb{P}_{j;\rho}^{\,i}(t)=\mathrm{\mathrm{Tr}}\left(\mathrm{vec}^{-1}\left[\left(\sum_{\alpha,\beta=1}^2\int_{\mathbb{R}}e^{-xt}\left[\mathcal{Q}_i^\alpha(x)\right]^*dW_{\alpha,\beta}(x)\mathcal{Q}_j^\beta(x)\right)\mathrm{vec}(\rho)\right]\right),\quad i,j\in\mathbb{Z}.
$$
The matrix-valued polynomials can be obtained by writing 
$$
\mathcal{Q}_i^\alpha(x)=BD_i^\alpha(x)B^{*},\quad D_i^\alpha(x)=\mbox{diag}(Q_i^{\lambda_1,\alpha}(x),Q_i^{\lambda_2,\alpha}(x),Q_i^{\lambda_3,\alpha}(x),Q_i^{\lambda_4,\alpha}(x)),\quad\alpha=1,2,\quad i\in\mathbb{Z},
$$
where the $\lambda_k, k=1,2,3,4,$ are the eigenvalues of $\ceil{T}$. Then, we have
\begin{equation*}\label{tqmc_calcZ}
\mathbb{P}_{j;\rho}^{\,i}(t)=\mathrm{\mathrm{Tr}}\left(\mathrm{vec}^{-1}\left[B\left(\sum_{\alpha,\beta=1}^2\int_{\mathbb{R}}e^{-xt}\left[D_i^\alpha(x)\right]^*D_j^\beta(x)\Psi_{\alpha,\beta}(x)dx\right)B^*\mathrm{vec}(\rho)\right]\right).
\end{equation*}
We recall that all these computations involve indexes $i,j\in\mathbb{Z}$. Again, Example \ref{Ex4.1} can be revisited in this setting, obtaining
$$
\mathbb{P}_{j;\rho}^{\, i}(t)=e^{-t}I_{i-j}(t),\quad i,j\in\mathbb{Z}.
$$
The diagrams in this situation locate exactly in between the diagrams of the reflecting and absorbing case of the half-integer line, see Figure \ref{fig3}. 
As for the site recurrence, we have now that $\mathbb{P}_{i;\rho}^{\, i}(t)=e^{-t}I_{0}(t)$, so, using again \eqref{mBf1LT} we get $\int_0^\infty\mathbb{P}_{i;\rho}^{\,i}(t)dt=\infty$ for every initial density $\rho$. Therefore we have that all vertices are recurrent.

\begin{remark}
Another way to derive formula \eqref{kmgZ} in this situation is via the Laplace transform applied to the difference-differential equations originated from $P'(t)=P(t)\mathcal{A}_\lambda, P(0)=I$, as it was done in \cite{tamon}. A second approach, regarding the folding trick of turning a walk on the integers into one on the half-integer line with a larger state space, is also possible here. For more details on the latter, we refer ther reader to \cite[Section 8]{dILL} or, more recently, \cite[Section 5]{New2}.
\end{remark}

\section{QMCs on the finite segment}\label{sec6}

In this case, and in a similar way as for the case of the half-line with reflecting barrier, by considering the spectrum of the map
$$\mathcal{L}=\Phi-I=\begin{bmatrix} T-I &  T &  &  &  \\
T & -I & T &  &  \\ 
 & T & -I & T &  \\
  &  & \ddots & \ddots & \ddots\\
 &  & & T & -I & T  \\
& &  & & T & T-I \end{bmatrix},$$
we are led to consider the generator
\begin{equation*}\label{Jfin}
\mathcal{A}_\lambda=\begin{bmatrix} \lambda-1 &  \lambda &  &  &  \\
{\lambda} & -1 & \lambda &  &  \\ 
 & {\lambda} & -1 & \lambda &  \\
  &  & \ddots & \ddots & \ddots\\
 &  & & {\lambda} & -1 & \lambda  \\
& &  & & \lambda & \lambda-1 \end{bmatrix}.
\end{equation*}
From the eigenvalue equation we have that $-xQ^\lambda(x)=Q^\lambda(x)\mathcal{A}_\lambda$, where $Q^\lambda(x)=(Q_0^\lambda(x), Q_1^\lambda(x),\ldots,Q_N^\lambda(x))$ with $Q_0^\lambda(x)=1$. The last component of the eigenvalue equation gives the relation $-xQ_N^\lambda(x)=\lambda Q_{N-1}^\lambda(x)+(\lambda-1)Q_N^\lambda(x)$. Since $\mbox{deg}(Q_n)=n$, the last relation will not hold in general \emph{unless} $x$ is an eigenvalue of $\mathcal{A}_\lambda$. Therefore, the eigenvalues of $\mathcal{A}_\lambda$ can be computed from the zeros of the polynomial
$$
R_{N+1}(x)=-(x+\lambda-1)Q_N^\lambda(x)-\lambda Q_{N-1}^\lambda(x)=(1-2\lambda-x)U_N\left(\frac{1-x}{2\lambda}\right),
$$
where $(Q_n^\lambda)_{n\geq0}$ are given by \eqref{pol1}. Clearly, these zeros are given by (see \eqref{zerosU})
\beq\label{formxdis}
x_{N+1,k}=1-2\lambda\cos\left(\frac{k\pi}{N+1}\right),\quad k=0,1,\ldots,N.
\eeq
The values of the jumps can also be computed. For that we can use 
$$
\alpha_k=\lim_{z\to x_{N+1,k}}(x_{N+1,k}-z)\left(\mathcal{A}_\lambda-zI\right)_{00}^{-1},\quad k=0,1,\ldots,N,
$$
or any other formula related with the corresponding orthogonal polynomials. After straightforward computations involving Chebychev polynomials, we have
\begin{equation*}\label{jumps}
\alpha_0=\frac{1}{N+1},\quad\alpha_k=\frac{1}{N+1}\left(1+\cos\left(\frac{k\pi}{N+1}\right)\right),\quad k=1,\ldots,N.
\end{equation*}
Therefore, the spectral measure is given by
$$
\boxed{\psi_\lambda(x)=\frac{1}{N+1}\delta_{x_{N+1,0}}(x)+\sum_{k=1}^N\frac{1}{N+1}\left(1+\cos\left(\frac{k\pi}{N+1}\right)\right)\delta_{x_{N+1,k}}(x)\;\;\;\text{(finite segment, reflecting)}}
$$
The orthogonal polynomials $(Q_n^\lambda)_n$ are the same as the ones in \eqref{pol1} but counting only the first $N+1$, i.e.
\begin{equation*}\label{pol1b}
Q_n^\lambda(x)=V_n\left(\frac{1-x}{2\lambda}\right),\quad n=0,1,\ldots,N,
\end{equation*}
where $(V_n)_n$ are the Chebychev polynomials of the third kind defined by \eqref{CP3}. In a similar way as we did in Section \ref{sec42}, we can give a solution of the differential equation $P'(t)=P(t)\mathcal{A}_\lambda, P(0)=I_{N+1},$ in terms of the Karlin-McGregor formula and also can be simplified after the change of variables \eqref{formxdis}. Indeed,
\begin{align*}
P_{ij}^\lambda(t)&=\sum_{k=0}^N e^{-x_{N+1,k}t}V_i\left(\frac{1-x_{N+1,k}}{2\lambda}\right)V_j\left(\frac{1-x_{N+1,k}}{2\lambda}\right)\alpha_k\\
&=\frac{e^{-t}}{N+1}\sum_{k=0}^N e^{2\lambda t\cos\left(\frac{k\pi}{N+1}\right)}\frac{\cos\left[(i+1/2)\frac{k\pi}{N+1}\right]\cos\left[(j+1/2)\frac{k\pi}{N+1}\right]}{\cos^2(\frac{k\pi}{2(N+1)})}\left(1+\cos\left(\frac{k\pi}{N+1}\right)\right)\\
&=\frac{2e^{-t}}{N+1}\sum_{k=0}^N e^{2\lambda t\cos\left(\frac{k\pi}{N+1}\right)}\cos\left[(i+1/2)\frac{k\pi}{N+1}\right]\cos\left[(j+1/2)\frac{k\pi}{N+1}\right]\\
&\frac{e^{-t}}{N+1}\sum_{k=0}^N e^{2\lambda t\cos\left(\frac{k\pi}{N+1}\right)}\left(\cos\left[(i-j)\frac{k\pi}{N+1}\right]+\cos\left[(i+j+1)\frac{k\pi}{N+1}\right]\right).
\end{align*}
Above we have used \eqref{CP3} and standard trigonometric formulas. We have not found explicit formulas for these discrete sums, but we guess that they may be related with some instances of discrete Bessel functions (see, for instance \cite{Boy,UW}).

\medskip

The absorbing case with generator
\begin{equation*}
\mathcal{A}_\lambda=\begin{bmatrix} -1 &  \lambda &  &  &  \\
{\lambda} & -1 & \lambda &  &  \\ 
 & {\lambda} & -1 & \lambda &  \\
  &  & \ddots & \ddots & \ddots\\
 &  & & {\lambda} & -1 & \lambda  \\
& &  & & \lambda & -1 \end{bmatrix},
\end{equation*}
is similar. It is possible to see that the spectral measure is given by
$$
\boxed{\psi_\lambda(x)=\frac{2}{N+2}\sum_{k=0}^N\sin^2\left(\frac{(k+1)\pi}{N+2}\right)\delta_{x_{N+1,k}}(x)\;\;\;\text{(finite segment, absorbing)}}
$$
where
$$
x_{N+1,k}=1-2\lambda\cos\left(\frac{(k+1)\pi}{N+1}\right),\quad k=0,1,\ldots,N.
$$
Also, we can combine absorbing and reflecting initial and final sites, getting similar results. It is possible to see, applying the same arguments as in Subsections \ref{sec32} and \ref{sec42}, that all vertices for the QMC on the finite segment with reflecting barriers are recurrent, while, if one of the barriers is absorbing, then all vertices are transient.

\bex Let
$$
B=\frac{1}{\sqrt{3}}\begin{bmatrix} 1 & 1 \\ 1 & 0\end{bmatrix},\;\;\;C=\frac{1}{\sqrt{3}}\begin{bmatrix} 0 & -1 \\ -1 & 1\end{bmatrix}\;\Rightarrow\; \ceil{T}=B\otimes B+C\otimes C=\frac{1}{3}\begin{bmatrix} 1 & 1 & 1 & 2\\ 1 & 0 & 2 & -1 \\ 1 & 2 & 0 & -1 \\ 2 & -1 & -1 & 1\end{bmatrix},$$
so $T(X)=B\cdot B^*+C\cdot C^*$ is a unital, trace-preserving CP map, but it is not a PQ-channel. Nevertheless, we are able to consider QMCs induced by them and obtain their measures. Note that $T$ is Hermitian and we have the unitary diagonalization
$$\ceil{T}=\hat{B}D_T\hat{B}^*,\;\;\;\hat{B}=\begin{bmatrix} 
\frac{1}{\sqrt{10}} & \frac{1}{\sqrt{2}} & -\frac{\sqrt{10}}{5} & 0 \\
\frac{\sqrt{10}}{5} & 0 & \frac{1}{\sqrt{10}} & -\frac{1}{\sqrt{2}} \\
\frac{\sqrt{10}}{5} & 0 & \frac{1}{\sqrt{10}} & \frac{1}{\sqrt{2}}\\
-\frac{1}{\sqrt{10}} & \frac{1}{\sqrt{2}} & \frac{\sqrt{10}}{5} & 0 \end{bmatrix},\;\;\;D_T=\mbox{diag}\Big(1,1-\frac{2}{3},-\frac{2}{3}\Big).$$
Let us consider the case that $N=4$, that is, we are considering 5 vertices. By taking $T/2$, we may examine the  QMC with absorbing barriers as above, so that 
$$
\int_{\mathbb{R}}e^{-xt}D_j^*(x)D_i(x)\Psi(x)dx=\begin{bmatrix}  P_{ij}^{\lambda_1}(t) & 0 & 0 & 0  \\ 0 & P_{ij}^{\lambda_2}(t) & 0 & 0 \\ 0 & 0 & P_{ij}^{\lambda_3}(t) & 0 \\ 0 & 0 & 0 & P_{ij}^{\lambda_4}(t)\end{bmatrix},
$$
where $P_{ij}^{\lambda_k}(t)$ is as above, for $\lambda_{1,2}=1/2$, $\lambda_{3,4}=-1/3$. Then, a calculation similar to the ones made in previous examples provides
$$
\mathbb{P}_{j;\rho}^{\, i}(t)=\mathrm{\mathrm{Tr}}\left(\mathrm{vec}^{-1}\left[\hat{B}\Big[\int_{\mathbb{R}}e^{-xt}D_i^*(x)D_j(x)\Psi(x)dx\Big]\hat{B}^{*}\mathrm{vec}(\rho)\right]\right)=P_{ij}^{\lambda_1}(t).$$
On the other hand, by considering a projection onto a specific state on vertex $i$, say $|\psi\rangle=[ 1\; 0]^T$, we obtain, given an initial density $\rho=\frac{1}{2}\begin{bmatrix} 1+X & Y+iZ \\ Y-iZ & 1-X\end{bmatrix}$, that
\begin{align*}
\mathbb{P}_{j;\rho}^{\, i;\gamma}(t)=\Big(P_{ij}^{\lambda_2}(t)+4P_{ij}^{\lambda_4}(t)\Big)\frac{1+X}{10}+
\Big(P_{ij}^{\lambda_2}(t)- P_{ij}^{\lambda_4}(t)\Big)\frac{Y}{5}+
\frac{P_{ij}^{\lambda_1}(t)}{2}-\frac{P_{ij}^{\lambda_2}(t)}{10}-\frac{2P_{ij}^{\lambda_4}(t)}{5},
\end{align*}
where $\gamma= |\psi\rangle\langle \psi|$. Also note that, as this is a semigroup acting on a finite number of vertices, one may calculate the matrix exponential of the generator and obtain the transition probabilities directly. The results obtained are in accordance with the above procedure, as expected.
\eex
\qee

\section{Outlook}\label{sec7}

1. {\bf Non-Hermitian and non-diagonalizable transitions.} There is a considerable range of examples for which the transition spectrum is not real. In \cite{dILL}, one can find evolutions for which one has matrix-valued measures that are not positive definite, so the Stieltjes transform is not available and one has to look for a distinct approach. Describing the associated spectral matrices for general homogeneous QMCs, i.e., induced by CP maps with non-Hermitian matrix representations is a problem which has not been completely solved, up to our knowledge.

\medskip

Similarly, we can also ask: what happens if the transition maps have matrix representations that a) are non-normal, or b) are non-diagonalizable? Examples of the former are given by the amplitude damping, which is a PQ-channel \cite{nielsen}, the channel defined by Attal et al. \cite[Section 4]{attal}, or \cite[Section 9]{dILL}; an example of the latter is given by the 1-qubit channel whose Kraus matrices are the rows of the Hadamard coin. It would be desirable to obtain, for such classes of examples, an approach which is somehow similar to what was presented in this article, in the sense that the relevant information is essentially determined by some scalar dynamics. If such a procedure exists at all, it seems that it must work around the absence of the Stieltjes transform on the line in some way. 

\medskip

Regarding the task of generating examples of  Lindblad operators $\mathcal{L}$ which are not of the form $\mathcal{L}=\Phi-I$, one can always begin with transitions and Hamiltonians $R_j^l$, $H_l$ of interest and then build $\mathcal{L}$ as in (\ref{lindbladg}). However, if one demands a positive definite measure, the criterion given by Theorem \ref{dette21} must be satisfied as well.

\medskip

2. {\bf Higher dimensional examples.} In the process of calculating probabilities for QMCs, it is noted that the results depend, in general, on the choice of the initial density. However, the oscillations obtained by choosing distinct densities do not seem to produce large changes in the numerical values. This is in part due to the fact that the internal degree is small, namely, a 2-dimensional vector space describing the spin of a electron. Then, a natural question is: what happens with the description of properties with larger internal degree of freedom? It is expected that oscillations in the probabilities will also occur, but a detailed study of such matters seems to be largely unexplored in the context of OQWs and QMCs. An interesting illustration regarding the possible evolution of densities under the action of discrete-time OQWs on the line can be seen in \cite[Section 4]{attal}, where densities of order 5 are studied under a particular example so that the limit behavior exhibits two Gaussians plus a Dirac soliton. It turns out that a central limit theorem is available for homogeneous OQWs with arbitrary (finite) internal degree of freedom \cite{carbone}, and it is believed that one such result is also valid for more general QMCs, but much remains to be done in this direction.

\medskip

{\bf Acknowledgements.} The authors would like to thank the anonymous reviewers for their comments, suggestions and corrections which have been really helpful to improve this manuscript. The work of MDI was partially supported by PAPIIT-DGAPA-UNAM grant IN106822 (M\'exico).

\appendix

\section{Chebychev polynomials and Bessel functions}\label{app1}
In this appendix we include some basic definitions and formulas related with Chebychev polynomials and Bessel functions. For more information the reader can consult \cite{EMOT, MHC, OB, Wat}.
\medskip

{\bf Chebychev polynomials of the first kind} $(T_n(x))_n$ can be defined as the unique polynomials satisfying
\begin{equation}\label{CP1}
T_n(\cos\theta)=\cos n\theta,\quad n\geq0,\quad x=\cos\theta\in[-1,1].
\end{equation}
They satisfy the three-term recurrence relation
\begin{equation}\label{CPTTRR}
T_0=1,\quad T_1(x)=x,\quad 2xT_n(x)=T_{n+1}(x)+T_{n-1}(x),\quad n\geq1,
\end{equation}
and are orthogonal with respect to the measure $\psi(x)=1/\sqrt{1-x^2}$ on $x\in[-1,1]$. From the definition \eqref{CP1} we have that the (real) zeros of $T_n(x)$ are given by
$$
x_{n,k}=\cos\left(\frac{(2k-1)\pi}{2n}\right),\quad k=1,\ldots,n.
$$
{\bf Chebychev polynomials of the second kind} $(U_n(x))_n$ can also be defined in a trigonometric way as
\begin{equation}\label{CP2}
U_n(\cos\theta)=\frac{\sin((n+1)\theta)}{\sin\theta},\quad n\geq0,\quad x=\cos\theta\in[-1,1].
\end{equation}
They satisfy the same three-term recurrence relation as the Chebychev polynomials of the first kind \eqref{CPTTRR}, but with $U_1(x)=2x$, and are orthogonal with respect to the measure $\psi(x)=\sqrt{1-x^2}$ on $x\in[-1,1]$. Again, from the definition \eqref{CP2} we have that the (real) zeros of $U_n$ are given by
\begin{equation}\label{zerosU}
x_{n,k}=\cos\left(\frac{k\pi}{n+1}\right),\quad k=1,\ldots,n.
\end{equation}
The following formula between Chebychev polynomials of the first and second kind will be useful in Section \ref{sec5}:
\begin{equation}\label{cp1cp2}
T_{n-m}(x)=T_n(x)U_m(x)-T_{n+1}(x)U_{m-1}(x),\quad n,m\geq0.
\end{equation}
Finally, the {\bf Chebychev polynomials of the third kind} $(V_n(x))_n$ can be defined as
\begin{equation}\label{CP3}
V_n(\cos\theta)=\frac{\cos((n+1/2)\theta)}{\cos(\theta/2)},\quad n\geq0,\quad x=\cos\theta\in[-1,1],
\end{equation}
and they satisfy the same three-term recurrence relation as the Chebychev polynomials of the first kind \eqref{CPTTRR}, but with $V_1(x)=2x-1$. They are orthogonal with respect to the measure $\psi(x)=\sqrt{1+x}/\sqrt{1-x}$ on $x\in[-1,1]$ and the zeros can also be explicitly computed. They are related with the Chebychev polynomials of the second kind through the formula
$$
V_n(x)=U_n(x)-U_{n-1}(x),\quad n\geq0.
$$
There is a fourth family of Chebychev polynomials but we will not be using them here (see \cite{MHC} for more details).
\medskip

The {\bf modified Bessel function of the first kind} can be defined as
\begin{equation}\label{mBf1}
I_\nu(x)=i^{-\nu}J_\nu(ix)=\sum_{m=0}^\infty\frac{1}{m!\Gamma(m+\nu+1)}\left(\frac{x}{2}\right)^{2m+\nu},
\end{equation}
where $J_\nu(x)$ is the Bessel function of the first kind. From formula (2) of \cite[p. 81]{EMOT} we have an integral representation for $I_\nu(x)$ when $\nu$ is an integer, given by (see \eqref{CP1})
\begin{equation}\label{mBf1IR}
I_n(x)=\frac{1}{\pi}\int_0^\pi e^{x\cos\theta}T_n(\cos\theta)d\theta=\frac{1}{\pi}\int_0^\pi e^{x\cos\theta}\cos(n\theta)d\theta,\quad n\in\mathbb{Z}.
\end{equation}
Recall that $I_{-n}(x)=I_n(x), n\in\mathbb{Z},$ as a consequence of $J_{-n}(x)=(-1)^nJ_n(x), n\in\mathbb{Z}$. Another useful formula is the Laplace transform of $I_\nu(\alpha x)$ for $\Re(\nu)>-1$, which can be found in \cite[Section 1.15]{OB}:
\begin{equation}\label{mBf1LT}
\int_0^{\infty}e^{-sx}I_\nu(\alpha x)dx=\frac{1}{\sqrt{s^2-\alpha^2}}\left(\frac{s-\sqrt{s^2-\alpha^2}}{\alpha}\right)^\nu,\quad\Re(s)>|\Re(\alpha)|.
\end{equation}
Finally, the modified Bessel function of the first kind has the following asymptotic (Hankel) expansion as $x\to\infty$:
\begin{equation}\label{mBf1HE}
I_\nu(x)\sim\frac{e^x}{\sqrt{2\pi x}}\sum_{k=0}^\infty(-1)^k\frac{a_k(\nu)}{x^k},\quad x\to\infty,\quad a_k(\nu)=\frac{\left(\frac{1}{2}-\nu\right)_k\left(\frac{1}{2}+\nu\right)_k}{(-2)^kk!},
\end{equation}
where $(b)_n=b(b+1)\cdots(b+n-1)$ denotes the usual Pochhammer symbol. From here we can clearly see that $\int_0^\infty e^{-x}I_\nu(x)$ is always divergent.

\end{document}